\documentclass[journal]{IEEEtran}
\usepackage{amsmath,amsfonts}
\usepackage{algorithmic}
\usepackage{algorithm}
\usepackage{array}
\usepackage[caption=false,font=normalsize,labelfont=sf,textfont=sf]{subfig}
\usepackage{textcomp}
\usepackage{stfloats}
\usepackage{url}
\usepackage{verbatim}
\usepackage{xcolor}
\usepackage{amsfonts}
\usepackage{amssymb}
\usepackage{bbm}
\usepackage{placeins}
\usepackage{float}
 \usepackage{booktabs} 
\usepackage{graphicx}
\usepackage{cite}
\usepackage{amsthm}
\newtheorem{Remark}{Remark}

\begin{document}
\bstctlcite{IEEEexample:BSTcontrol}

\title{Hybrid Six-Level Rydberg Atomic Quantum Receiver for Multi-Band Wireless Communications}

\author{Lahiru Shyamal,~\IEEEmembership{Student Member,~IEEE,}
Harini Hapuarachchi,\\
Saman Atapattu,~\IEEEmembership{Senior~Member,~IEEE,}
and Jared H. Cole
\vspace{-0.5cm}}


\maketitle

\begin{abstract}
Rydberg atomic quantum receivers (RAQRs) have recently emerged as a
promising technology for radio-frequency (RF) reception by directly
transducing incident RF fields into optical signals. Existing receiver
architectures, however, exploit only subsets of the dipole-allowed
transitions within a given atomic manifold, limiting the number of
simultaneously accessible RF channels. In this paper, a hybrid six-level
Rydberg atomic quantum receiver (H-RAQR) is proposed by integrating
parallel and cascaded RF coupling pathways within a single vapor-cell
receiver. A communication-oriented analytical framework is developed by
deriving a closed-form steady-state atom--field interaction model and
establishing an equivalent baseband signal representation. The
achievable ergodic sum rate is analyzed, and a resource-efficiency
metric is introduced to quantify throughput per unit optical receiver
resource. The analytical model is validated against full Lindblad
master-equation simulations over its identified operating region.
Numerical results show that the proposed H-RAQR supports four
simultaneous RF channels within a single atomic system, achieves higher
ergodic sum rate than conventional parallel Rydberg state (PRS) and
cascade Rydberg state (CRS) receivers, and provides about $29\%$ higher
resource efficiency than a combined PRS--CRS deployment with equivalent
four-band coverage. The proposed framework provides a scalable foundation
for multi-band atomic wireless receivers.
\end{abstract}

\begin{IEEEkeywords}
Rydberg atomic receiver, multi-band reception, wireless communications,
communication theory, electromagnetically induced transparency (EIT).
\end{IEEEkeywords}

\vspace{-2mm}

\section{Introduction}

Atomic systems have emerged as a promising platform for radio-frequency
(RF) reception, offering capabilities beyond conventional electronic
receivers. In particular, Rydberg Atomic Quantum Receivers (RAQRs)
exploit the exceptionally large electric dipole moments and strong
electric-field sensitivity of highly excited Rydberg atoms to directly
interact with incident RF fields over a broad frequency range. These properties have enabled applications in RF sensing, precision
metrology, quantum sensing, wireless
communications, satellite
communications ~\cite{10972179,11124471},
underwater communication~\cite{shyamal2026rydberg}, and radar
sensing~\cite{watterson2025imagingradarusingrydberg}.
Using electromagnetically induced transparency (EIT), RAQRs coherently
transduce incident RF fields into optical probe signals, enabling
all-optical RF detection without conventional electronic front-end
circuitry~\cite{10845209,11126890,Atapattu26qcnc}. From a communication receiver
perspective, RAQRs provide a fundamentally different approach to
multi-band reception: a single vapor-cell platform can simultaneously
receive signals spanning sub-6~GHz to terahertz (THz) frequencies,
potentially eliminating the need for dedicated RF chains for individual
frequency bands~\cite{10972179,Holloway2017}. Reported sensitivities are
on the order of $\mathrm{nV/cm/\sqrt{Hz}}$, with theoretical limits
approaching $\mathrm{pV/cm/\sqrt{Hz}}$ under the standard quantum
limit~\cite{Jing2020,10972179}, making RAQRs attractive for weak-signal
RF reception~\cite{Liu2023,Holloway2022}.

\subsection{Related Works}

Early Rydberg atomic receivers were based on four-level ladder-type
configurations, where probe and coupling lasers establish an EIT window
while a single RF field couples two high-lying Rydberg
states~\cite{10972179}. These architectures demonstrated coherent
reception of digitally modulated signals, including BPSK, QPSK, and QAM,
establishing RAQRs as phase-sensitive communication
receivers~\cite{8778739}. Subsequent heterodyne receiver architectures
incorporating RF local oscillators (LOs) improved sensitivity and enabled
frequency down-conversion~\cite{newTAP2}. More recently, Rydberg
receivers have been extended to enhanced-sensitivity
architectures~\cite{10803071}, broadband and array-based
receivers~\cite{10845209,11126890}, and communication-oriented system
models. In particular, Gong \textit{et al.}~\cite{Gong2025AnEB}
developed an equivalent baseband communication model, while Cui
\textit{et al.}~\cite{10845209} proposed a Rydberg-based MIMO receiver
architecture. To further increase multi-band reception capability, two
receiver architectures have recently emerged: \emph{Parallel Rydberg
States} (PRS) and \emph{Cascade Rydberg States} (CRS).

PRS architectures exploit multiple RF-accessible Rydberg transitions
through parallel excitation pathways, enabling simultaneous reception
across multiple frequency bands within a single vapor
cell. Experimental demonstrations include simultaneous demodulation of
five RF signals spanning $1.7$--$116$~GHz~\cite{PhysRevApplied.19.014025},
while communication-oriented extensions include a multi-user reception
framework~\cite{11124471} and a generalized $N$-level multi-band receiver
model spanning MHz-to-THz frequencies~\cite{cui2026rydbergatomicreceiversmultiband}.
In contrast, CRS architectures employ cascaded excitation pathways
between sequential Rydberg states to support wideband RF sensing and
coherent signal processing. Representative examples include a five-level
receiver based on cascaded EIT-Autler-Townes
splitting~\cite{JIN2022128603} and a numerical framework for general
$N$-level cascade systems~\cite{Bussey2024}. While PRS and CRS
architectures significantly extend the capabilities of conventional
four-level receivers, they exploit complementary subsets of the
available dipole-allowed transitions within a given atomic manifold.

\subsection{Research Gap, Motivation, and Contributions}

Modern wireless communication systems increasingly require simultaneous
multi-band reception across heterogeneous RF spectrum bands.
Conventional multi-band receivers achieve this using multiple parallel
RF front-ends, resulting in increased hardware complexity and power
consumption. Rydberg atomic receivers provide an alternative by
transducing multiple RF signals within a single vapor cell through
distinct atomic transitions. Existing PRS and
CRS architectures, however, exploit only
complementary subsets of the dipole-allowed transitions available within
a given atomic manifold
~\cite{PhysRevApplied.19.014025,11124471,
cui2026rydbergatomicreceiversmultiband,JIN2022128603,Bussey2024}.
Consequently, the number of simultaneously accessible RF channels and
the achievable communication throughput remain fundamentally limited.

Motivated by this observation, we propose a \emph{Hybrid Rydberg Atomic
Quantum Receiver} (H-RAQR) that unifies PRS and CRS coupling pathways
within a single atomic system to exploit the full dipole-allowed
connectivity of the Rydberg manifold. More importantly, this work
develops a communication-theoretic framework for hybrid Rydberg
receivers by establishing the relationship between atom--field
interactions, the equivalent baseband communication model, and
receiver-level performance metrics. This framework enables systematic
analysis and optimization of hybrid atomic receiver architectures prior
to experimental implementation. 
The main contributions of this work are summarized as follows:

\begin{itemize}

\item A hybrid six-level Rydberg atomic quantum receiver (H-RAQR) is
proposed by integrating parallel and cascaded RF coupling pathways,
enabling simultaneous multi-band RF reception through full exploitation
of the dipole-allowed transitions within a single atomic manifold.

\item A communication-oriented analytical framework is developed by
deriving the first closed-form steady-state probe-coherence expression
for the proposed hybrid architecture, establishing an equivalent
baseband communication signal model, and validating the analytical model
against full Lindblad master-equation simulations.

\item Communication performance is characterized through ergodic
sum-rate analysis, and a resource-efficiency metric is introduced to
quantify throughput per unit optical receiver resource, demonstrating
the advantages of the proposed H-RAQR over conventional PRS and CRS
receiver architectures.

\end{itemize}

The proposed framework is evaluated using full Lindblad master-equation simulations with experimentally reported system parameters, enabling
systematic communication-level performance analysis prior to hardware
implementation.


{\it Notation:}
Matrices and operators are denoted by bold symbols (e.g., $\mathbf{A}$), while physical vector quantities are denoted by overhead arrows (e.g., $\vec{E}$). Quantum states are represented using Dirac notation, with ket and bra vectors denoted by $|\psi\rangle$ and $\langle\psi|$, respectively. The commutator and anticommutator are defined as $[\mathbf{A},\mathbf{B}] = \mathbf{A}\mathbf{B} - \mathbf{B}\mathbf{A}$ and $\{\mathbf{A},\mathbf{B}\} = \mathbf{A}\mathbf{B} + \mathbf{B}\mathbf{A}$. Hermitian conjugation and complex conjugation are denoted by $(\cdot)^{\dagger}$ and $(\cdot)^*$, respectively. The imaginary unit is $\mathbbm{i}=\sqrt{-1}$, the identity operator is $\mathbf{I}$, and the trace is $\mathrm{Tr}(\mathbf{A})$.


\section{System and Signal Model}\label{sec:SystemModel}
This section introduces the RF sensing environment, the receiver's 
sensing objective, and the proposed hybrid Rydberg atomic coupling 
architecture for simultaneous multi-band RF reception within a 
single vapor-cell platform.

\subsection{Network Model and Sensing Objective}

\begin{figure}[t]
\centering
\includegraphics[width=0.9\linewidth]{ 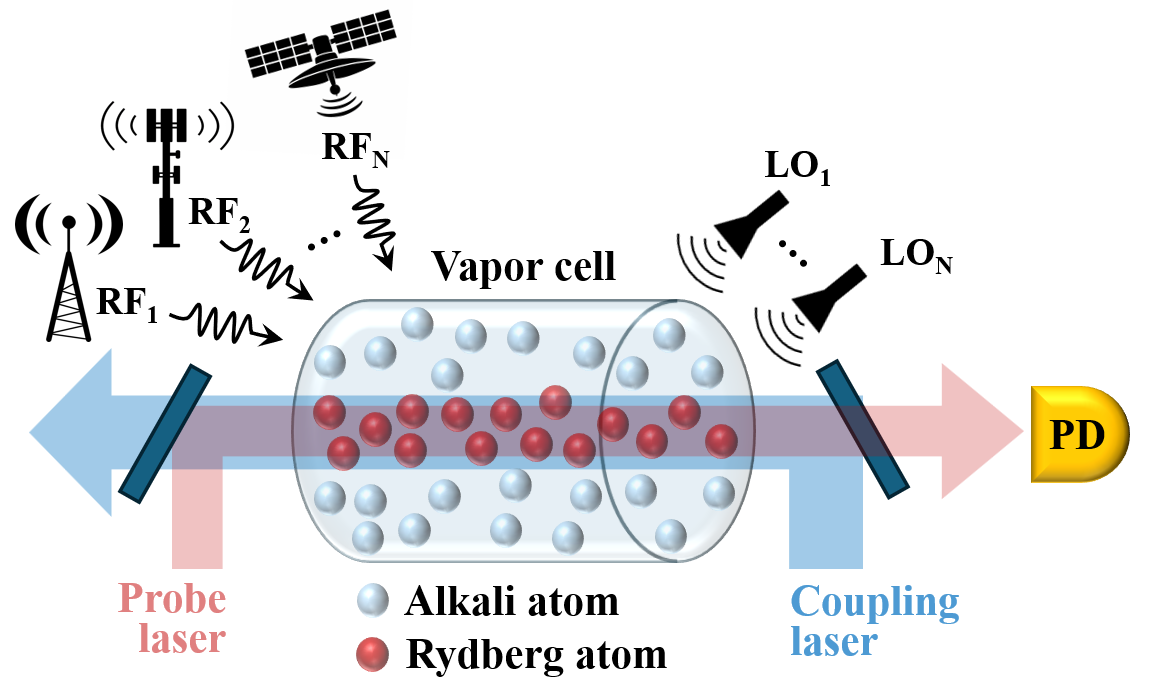} \caption{Rydberg atoms in a vapor cell interacting with probe and coupling 
lasers, RF signals to be detected, and local oscillators (LOs).
\label{fig:system_model}}
\end{figure}

We consider a wireless RF environment consisting of $N$ independent
radio-frequency (RF) transmitters operating at distinct carrier
frequencies. These transmitters may represent communication nodes,
sensing targets, or external emitters distributed within the coverage
region of the receiver. From an RF sensing and antenna perspective, the
receiver observes the superposition of incident electromagnetic fields
radiated by these transmitters. The objective of the receiver is to
simultaneously identify the individual RF carriers and extract their
modulation information using a single sensing platform.

Let the $n$-th transmitter radiate a narrowband RF signal centered at
angular frequency $\omega_{RF,n}$. The transmitted complex baseband
signal is denoted by $x_n(t) \in \mathbb{C}$ with average power
$P_{T,n}$. After propagating through a wireless channel with complex
coefficient $h_n \in \mathbb{C}$, the resulting electric field impinging
on the receiver is
\begin{equation}
E_{\mathrm{RF},n}(t)
=
\mathrm{Re}
\left\{
\sqrt{P_{T,n}}\, h_n\, x_n(t)\,
e^{\mathbbm{i}\omega_{RF,n} t}
\right\}.
\label{eq:RF_component}
\end{equation}

Assuming the RF carriers occupy distinct spectral bands, the total incident RF  field  is \(E_{\mathrm{RF}}(t)
=
\sum_{n=1}^{N}
E_{\mathrm{RF},n}(t).\)  The receiver, illustrated in Fig.~\ref{fig:system_model}, is a 
\emph{Rydberg Atomic Quantum Receiver} (RAQR) employing a vapor 
cell containing an ensemble of alkali atoms that interact coherently 
with optical and RF electromagnetic fields. Incident RF electric fields resonantly couple Rydberg states 
through electric-dipole transitions, imprinting RF information 
onto the atomic coherence. This atomic response is subsequently read out optically by 
monitoring the transmitted intensity of a probe laser through a 
single photodetector (PD). Unlike classical receivers, which require 
multiple antennas and dedicated RF front-end circuits for each frequency 
band, a single vapor-cell platform can simultaneously transduce signals 
spanning sub-6~GHz to terahertz (THz) frequencies, offering a unified 
hardware solution for diverse sensing and communication applications. The sensing objective is therefore to identify 
the carrier frequencies $\{\omega_n\}_{n=1}^{N}$ and recover the 
baseband signals $\{x_n(t)\}$ simultaneously.

\begin{figure*}[t]
\centering
\includegraphics[width=0.8\linewidth]{ 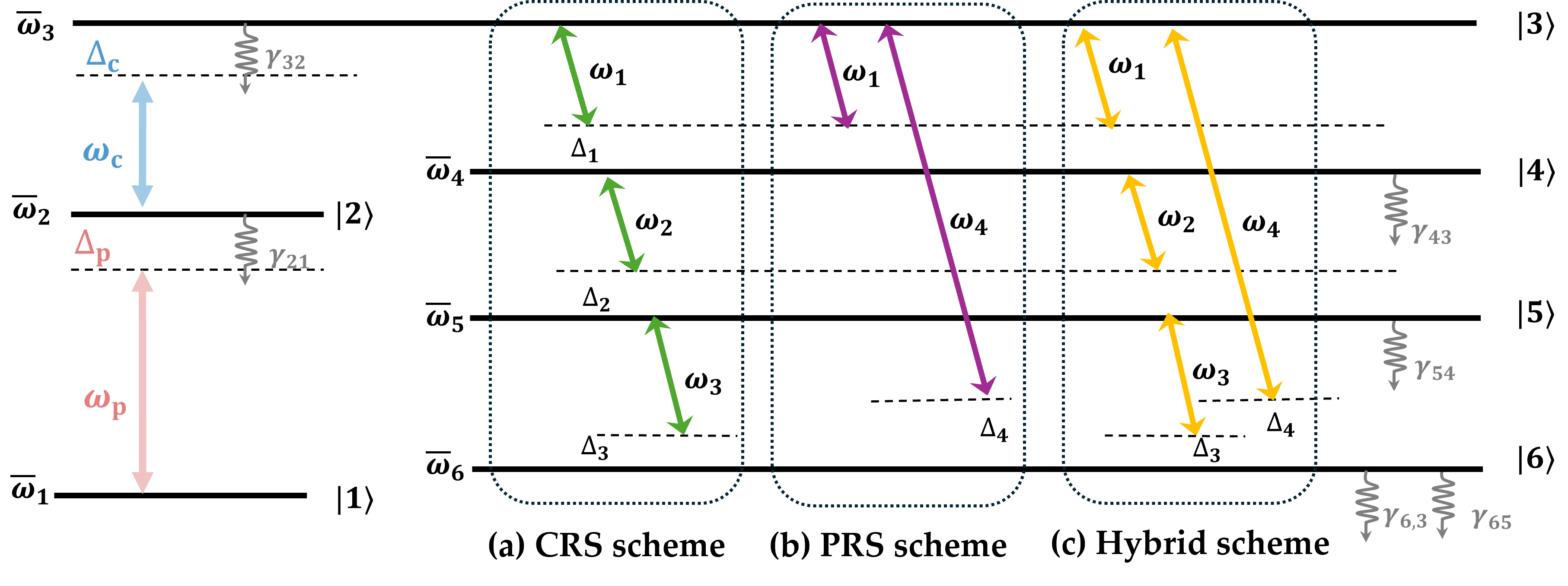} \caption{Six-level Rydberg coupling schemes: $\omega_n$ and $\Delta_n$ are the field frequency and detuning, $\bar{\omega}_j$ the eigenfrequency of level $j$, and $\gamma_{i,j}$ the decay rate from $i$ to $j$. (a) CRS configuration (b) PRS configurations. (c) H-RAQR configurations.
\label{fig:level_schemes}}
\end{figure*}

To enable simultaneous interaction with multiple RF carriers, we introduce a hybrid Rydberg coupling architecture that combines sequential (CRS) and parallel (PRS) excitation pathways within a single atomic manifold. This architecture allows a single vapor-cell receiver equipped with one photodetector to perform multi-frequency RF sensing and signal extraction. The atomic structure and interaction mechanisms underlying the proposed hybrid RAQR are described next.

\subsection{Proposed Hybrid Rydberg Atomic Architecture}

Let the atomic manifold contain $K$ levels $\{|1\rangle,\dots,|K\rangle\}$, where $|1\rangle$, $|2\rangle$, $|3\rangle$ denote the ground, intermediate, and initial Rydberg states, and $\{|4\rangle,\dots,|K\rangle\}$ form the RF-driven manifold. RF transitions within this manifold are governed by the 
electric-dipole selection rule, which requires opposite-parity 
states~\cite{bransden2003physics,haken2005physics}, constraining 
the set of accessible coupling pathways as stated in the following 
remark.

\begin{Remark}[Electric-dipole selection rule]\label{re:parity}
A transition $|i\rangle \leftrightarrow |j\rangle$ is dipole-allowed 
only if $\langle i|\vec{\boldsymbol{\mu}}|j\rangle \neq 0$, where 
$\vec{\boldsymbol{\mu}}$ is the atomic dipole operator. Since the 
dipole operator is odd under spatial inversion, this requires the 
two states to have opposite parity, i.e., 
$\pi_i\pi_j = -1$~\cite{bransden2003physics,haken2005physics}, 
where $\pi_k \in \{+1,-1\}$ denotes the parity of state $|k\rangle$.  
Consequently, the set of feasible RF coupling pathways in a 
multi-level Rydberg manifold is constrained by this parity condition.
\end{Remark}

Within the same $K$-level manifold, the three coupling architectures 
differ in how they exploit the dipole-allowed transitions. In a CRS architecture, RF fields couple adjacent states sequentially ($|3\rangle\leftrightarrow|4\rangle\leftrightarrow\cdots\leftrightarrow|K\rangle$), supporting $N_{\mathrm{CRS}}(K)=K-3$ transitions but with limited multi-band flexibility. In a PRS architecture, multiple RF fields couple $|3\rangle$ directly to even-offset levels ($|4\rangle,|6\rangle,\dots$), yielding $N_{\mathrm{PRS}}(K)=\lfloor(K-2)/2\rfloor$ transitions but leaving intermediate couplings unused.

To overcome these limitations, the proposed H-RAQR combines CRS-type sequential links with additional PRS-type non-adjacent branches from $|3\rangle$, wherever permitted by the dipole selection rule. The total number of accessible RF transitions is
\(
N_{\mathrm{hyb}}(K) = (K-3)+\left\lfloor\frac{K-4}{2}\right\rfloor,
\label{eq:no_of_bands}
\) providing a strictly richer RF sensing manifold within the same atomic structure.
\begin{Remark}[Minimal hybrid manifold]
\label{re:minlevels}
For $K=5$, hybrid operation requires both the ladder pathway 
$|3\rangle \leftrightarrow |4\rangle \leftrightarrow |5\rangle$ 
and the non-adjacent branch $|3\rangle \leftrightarrow |5\rangle$, 
forming a closed triangular loop. However, by 
Remark~\ref{re:parity}, the parity constraint $\pi_i\pi_j=-1$ 
cannot be simultaneously satisfied for all three transitions in 
this loop, meaning at least one transition is always dipole-forbidden. 
Hence a five-level manifold cannot realize hybrid coupling, and 
the smallest feasible configuration occurs at $K=6$.
\end{Remark}
As established in Remark~\ref{re:minlevels}, the 
minimal feasible hybrid configuration is $K=6$, and we consider 
this six-level realization in this work. This configuration captures the essential coexistence of cascaded  transitions and non-adjacent RF couplings while remaining analytically tractable. 
Hybrid architectures with larger manifolds ($K>6$) can be constructed following the same principle, but their quantum dynamics become increasingly complex due to the rapid growth of coupling pathways and density-matrix elements. 
The electromagnetic excitation model for the six-level system is developed next.

\section{Analytical Framework of Six-Level H-RAQR}
\label{Sec:AnalyFrameSix-Level}

To analyze the sensing behavior of the proposed H-RAQR, we develop 
a quantum dynamical model for the minimal six-level hybrid system. The interaction between the 
optical fields, RF fields, and the atomic ensemble is governed by 
the electric–dipole interaction Hamiltonian, while the resulting 
system dynamics are described using the density-matrix formalism 
for open quantum systems. The three six-level coupling schemes for 
$K=6$ are illustrated in Fig.~\ref{fig:level_schemes}.

From Section~\ref{eq:no_of_bands}, the six-level H-RAQR supports 
$N_{\mathrm{hyb}}(6)=4$ simultaneous RF channels, compared to 
$N_{\mathrm{CRS}}(6)=3$ for the CRS transition set 
$\{\omega_1,\omega_2,\omega_3\}$ and $N_{\mathrm{PRS}}(6)=2$ 
for the PRS transition set $\{\omega_1,\omega_4\}$, as shown in 
Figs.~\ref{fig:level_schemes}(a) and (b), respectively. . The four dipole-allowed RF transitions of the H-RAQR 
(Fig.~\ref{fig:level_schemes}(c)) are
$\mathcal{T}_1{:}\ |3\rangle \leftrightarrow |4\rangle$,
$\mathcal{T}_2{:}\ |4\rangle \leftrightarrow |5\rangle$,
$\mathcal{T}_3{:}\ |5\rangle \leftrightarrow |6\rangle$, and
$\mathcal{T}_4{:}\ |3\rangle \leftrightarrow |6\rangle$,
each characterized by the parameter tuple
\begin{equation}
\mathcal{T}_n = (\Omega_n,\,\omega_n,\,\Delta_n,\,\mu_n),
\label{eq:transition_sets}
\end{equation}
where $\Omega_n = \mu_n E_n/\hbar$ is Rabi frequency,
$\Delta_n$ is detuning,
$\mu_n=\langle i|\vec{\boldsymbol{\mu}}|j\rangle$ is
electric-dipole matrix element, $E_n$ is RF field envelope,
and $\gamma_{i,j}$ denotes spontaneous decay rate from
$|i\rangle$ to $|j\rangle$. In the following subsections, we derive
electromagnetic excitation model and density-matrix evolution
equations governing the six-level H-RAQR dynamics.


\subsection{Electromagnetic Field Interaction}

The operation of a Rydberg atomic quantum receiver relies on the coherent interaction between an atomic ensemble and optical as well as RF electromagnetic fields. As established in early works \cite{10845209,8778739}, the central sensing element is a vapor cell containing alkali atoms such as rubidium (Rb) or cesium (Cs). A probe laser with Rabi frequency $\Omega_P$ drives the transition
\(
|1\rangle \leftrightarrow |2\rangle,
\)
while a counter–propagating coupling laser with Rabi frequency $\Omega_C$ excites the atoms to the Rydberg state
\(
|2\rangle \leftrightarrow |3\rangle .
\)
Following standard conventions \cite{10845209}, the optical Rabi frequencies are
\begin{equation}
\Omega_P=\mu_{12}A_P/\hbar, 
\quad \text{and} \quad
\Omega_C=\mu_{23}A_C/\hbar,
\label{eq:optical_rabi}
\end{equation}
where $A_P$ and $A_C$ denote the electric–field envelopes of the probe and coupling lasers, respectively, and $\mu_{ij}$ represents the electric–dipole matrix element of the transition $|i\rangle\leftrightarrow|j\rangle$.

Once the atoms are prepared in the Rydberg state $|3\rangle$, the incident RF fields interact with the Rydberg manifold through dipole-allowed transitions indexed by $n=1,\dots,4$ as defined in \eqref{eq:transition_sets}. Each $n$-th RF channel, therefore, drives a transition with angular frequency $\omega_n$.

The received RF signal associated with the $n$-th transmitter, modeled in \eqref{eq:RF_component} can be expressed in terms of its instantaneous amplitude and phase as $E_{\mathrm{RF},n}(t) = A_{\mathrm{RF},n}\cos(\omega_{\mathrm{RF},n} t + \phi_{\mathrm{RF},n})$, where $A_{\mathrm{RF},n} = \sqrt{P_{T,n}} |h_n| |x_n|$ and $\phi_{\mathrm{RF},n} = \angle h_n + \angle x_n$. The LO field and the received RF signal related to the $n$-th transition are thus
\begin{align}
E_{\mathrm{LO},n}(t) &= A_{\mathrm{LO},n}\cos(\omega_{\mathrm{LO},n} t + \phi_{\mathrm{LO},n}), \notag \\
E_{\mathrm{RF},n}(t) &= A_{\mathrm{RF},n}\cos(\omega_{\mathrm{RF},n} t + \phi_{\mathrm{RF},n}),
\label{eq:physical_RF_field}
\end{align}
where $A_{\mathrm{LO},n}$ and $A_{\mathrm{RF},n}$ denote the corresponding field amplitudes. The LO and RF fields are assumed to be polarized parallel to 
the coupling laser, consistent with typical Rydberg receiver 
implementations~\cite{cui2026rydbergatomicreceiversmultiband}.
Total RF electric field interacting with the atoms is therefore
\(E_n(t)=E_{\mathrm{LO},n}(t)+E_{\mathrm{RF},n}(t)\).
Under the heterodyne condition $A_{\mathrm{LO},n}\gg A_{\mathrm{RF},n}$, the combined field can be approximated as \cite{Gong2025AnEB} 
\begin{equation}
E_n(t)\approx
\left[A_{\mathrm{LO},n}
+
A_{\mathrm{RF},n}\cos(\delta\omega_nt+\delta\phi_n)
\right]
\cos(\omega_{n}t+\phi_{\mathrm{LO},n}),
\label{eq:heterodyne_field}
\end{equation}
where $\delta\omega_n=\omega_{\mathrm{RF},n}-\omega_{\mathrm{LO},n}$ and $\delta\phi_n=\phi_{\mathrm{RF},n}-\phi_{\mathrm{LO},n}$. In subsequent sections, $\omega_n \triangleq \omega_{\mathrm{LO},n}$ 
denotes the carrier frequency of the $n$-th RF-driven transition.
The corresponding LO and RF signal Rabi frequencies are 	\cite{ScullyZubairy1997}
\begin{equation}
\Omega_{\mathrm{LO},n}=\frac{\mu_n A_{\mathrm{LO},n}}{\hbar}e^{\mathbbm{i}\phi_{\mathrm{LO},n}};
\,\,
\Omega_{\mathrm{RF},n}=\frac{\mu_n A_{\mathrm{RF},n}}{\hbar}e^{\mathbbm{i}\phi_{\mathrm{RF},n}} .
\label{eq:Rabi_RF}
\end{equation} 
Since the LO phase is controllable, we set $\phi_{\mathrm{LO},n}=0$ such that $\Omega_{\mathrm{LO},n}\in\mathbb{R}^+$. Under the condition $\Omega_{\mathrm{LO},n}\gg\Omega_{\mathrm{RF},n}$, the effective time-dependent Rabi frequency becomes \cite{cui2026rydbergatomicreceiversmultiband}
\begin{equation}
\Omega_n(t)
\approx
\Omega_{\mathrm{LO},n}
+
\frac{\mu_n}{\hbar}A_{\mathrm{RF},n}
\cos(\delta\omega_nt+\delta\phi_n),
\label{eq:rabi_heterodyne_second}
\end{equation}
which shows that received RF signal is down-converted to a low-frequency modulation of Rabi frequency. This forms the basis of atomic heterodyne detection in the proposed H-RAQR.


\section{Analytical Model for the proposed HRAQR}

Building on the coupling architecture introduced in Section~\ref{Sec:AnalyFrameSix-Level}, we 
develop the quantum dynamical model of the six-level H-RAQR, derive 
the closed-form steady-state probe coherence, and assess the validity and limitations of the resulting analytical model.

\subsection{Open Quantum System Model}

RAQRs operate in realistic vapor-cell environments, where the atomic ensemble is continuously driven by optical and RF fields while simultaneously interacting with its surrounding environment. 
As a result, the system cannot be treated as an isolated quantum system. 
Instead, dissipative processes such as spontaneous emission play a crucial role in determining the receiver dynamics and detection performance. To accurately capture both coherent atom–field interactions and irreversible environmental effects, we model the six-level Rydberg atomic receiver as an open quantum system.

Following the Schrödinger picture in \cite{allen2012optical}, the driven six-level atomic system shown in Fig.~\ref{fig:level_schemes}(c), together with the applied optical and RF fields, is described by the time-dependent Hamiltonian $\tilde{\boldsymbol{H}}(t)$
\begin{equation}
\tilde{\boldsymbol H}(t)=\boldsymbol H_0+\boldsymbol H_I(t),
\label{eq:H_lab_full} 
\end{equation}
where the bare atomic Hamiltonian is
$\boldsymbol H_0=\sum_{j=1}^{6}\hbar\bar{\omega}_j|j\rangle\langle j|$,
and the atom-field interaction is governed by the electric-dipole Hamiltonian
$\boldsymbol H_I(t)=-\vec{\boldsymbol{\mu}} \cdot\vec E(t)$. For the six-level coupling scheme in 
Fig.~\ref{fig:level_schemes}(c), the interaction Hamiltonian is 
\begin{align}
\boldsymbol H_I(t)
= &-\frac{\hbar}{2}\Big[
\Omega_P e^{-\mathbbm{i}\omega_P t}|1\rangle\langle2|
+\Omega_C e^{-\mathbbm{i}\omega_C t}|2\rangle\langle3|
\nonumber\\
&+\Omega_1 e^{-\mathbbm{i}\omega_1 t}|3\rangle\langle4|
+\Omega_2 e^{-\mathbbm{i}\omega_2 t}|4\rangle\langle5|
\nonumber\\
&+\Omega_3 e^{-\mathbbm{i}\omega_3 t}|5\rangle\langle6|
+\Omega_4 e^{-\mathbbm{i}\omega_4 t}|3\rangle\langle6|
+\mathrm{H.c.}
\Big],
\label{eq:ele_H}
\end{align}
where $\Omega_P$ and $\Omega_C$ are the optical Rabi frequencies defined in \eqref{eq:optical_rabi}, and $\Omega_n$ denotes the complex Rabi frequency associated with the $n$th RF-driven transition. For compactness, $\mathrm{H.c.}$ denotes Hermitian conjugate of the preceding interaction terms.

 The Hamiltonian in ~\eqref{eq:H_lab_full} contains rapidly oscillating optical and RF frequency components, which make the physically relevant slow dynamics challenging to solve directly. 
To obtain a tractable description, we transform the Hamiltonian into an interaction picture defined by the cumulative drive frequencies along the excitation ladder \cite{ScullyZubairy1997,Paing_2025}
\begin{equation}
    \boldsymbol{H}(t) = \boldsymbol{U}(t) \tilde{\boldsymbol{H}}(t) \boldsymbol{U}^\dagger(t) + \mathbbm{i}\hbar \frac{d\boldsymbol{U}(t)}{dt} \boldsymbol{U}^\dagger(t)
    \label{RWA_Transform}
\end{equation}
where $\boldsymbol{U}(t)=\operatorname{diag}\!\left(e^{\mathbbm{i}
\Phi_k t}\right)_{k=1}^{6}$, with  $\Phi_1\triangleq0$, $\Phi_2\triangleq\omega_P$, $\Phi_3\triangleq\omega_P+\omega_C$, and $\Phi_{k}\triangleq\Phi_3+\sum_{n=1}^{k-3}\omega_n$ for $k\in\{4,5,6\}$. After applying the transformation and neglecting rapidly oscillating terms using the rotating wave approximation (RWA), the effective rotating-frame Hamiltonian takes the form given in \eqref{eq:H6_general} (top of this page), where $\Delta_P=(\bar{\omega}_2-\bar{\omega}_1)-\omega_P$, $\Delta_C=(\bar{\omega}_3-\bar{\omega}_2)-\omega_C$, and $\Delta_n=(\bar{\omega}_{n+3}-\bar{\omega}_{n+2})-\omega_n$ for $n\in\{1,2,3\}$, $\Delta_4=(\bar{\omega}_6-\bar{\omega}_3)-\omega_4$ are the optical and RF detunings, $\Delta_{\Sigma}^{(m)}\triangleq\Delta_P+\Delta_C+\sum_{n=1}^{m}\Delta_n$ for $m=1,2,3$ are the cumulative detunings. And the residual phase mismatch in the non-adjacent hybrid branch is governed by the closed-loop RF detuning \(\delta\triangleq\Delta_4-\sum_{n=1}^{3}\Delta_n.\)

\begin{figure*}[t]
\begin{equation}
\boldsymbol H(t)=\hbar
\begin{pmatrix}
0 & \Omega_P/2 & 0 & 0 & 0 & 0 \\
\Omega_P^{*}/2 & -\Delta_P & \Omega_C/2 & 0 & 0 & 0 \\
0 & \Omega_C^{*}/2 & -(\Delta_P+\Delta_C) & \Omega_1/2 & 0 & \Omega_4 e^{-\mathbbm{i}\delta t}/2 \\
0 & 0 & \Omega_1^{*}/2 & -\Delta_{\Sigma}^{(1)} & \Omega_2/2 & 0 \\
0 & 0 & 0 & \Omega_2^{*}/2 & -\Delta_{\Sigma}^{(2)} & \Omega_3/2 \\
0 & 0 & \Omega_4^{*} e^{\mathbbm{i}\delta t}/2 & 0 & \Omega_3^{*}/2 & -\Delta_{\Sigma}^{(3)}
\end{pmatrix},
\label{eq:H6_general}
\end{equation}
\end{figure*}

\begin{Remark}[Closed-Loop RF Detuning]
The parameter $\delta$ quantifies the frequency mismatch accumulated 
around the RF loop
$|3\rangle \leftrightarrow |4\rangle \leftrightarrow |5\rangle 
\leftrightarrow |6\rangle \leftrightarrow |3\rangle$.
When $\delta\neq 0$, the direct hybrid branch 
$|3\rangle\leftrightarrow|6\rangle$ retains the residual phase factor 
$e^{-\mathbbm{i}\delta t}$, which acts as a persistent coherent drive 
on the coherence $\rho_{6,3}$, sustaining permanent oscillations in 
the population and coherence dynamics. This behavior arises because 
when $\delta\neq 0$, no stationary rotating frame exists in which all 
interaction terms simultaneously become time independent. We therefore restrict attention to the resonant 
regime $\delta = 0$, under which the Hamiltonian becomes time 
independent and a well-defined steady-state solution exists.
\end{Remark}

Following prior works on RAQRs \cite{cui2026rydbergatomicreceiversmultiband,Jing2020},
we consider the resonant operating regime in which the probe laser, coupling laser, and all RF local oscillators are tuned to their respective atomic transitions, and the drive phases are taken to be zero without loss of generality. Under these assumptions
\begin{equation}
\Delta_P=\Delta_C=\Delta_n=\delta=0,
\qquad n\in\{1,2,3,4\},
\label{eq:resonant_cond}
\end{equation}
and all Rabi frequencies may be chosen as real-valued. The rotating-frame Hamiltonian in \eqref{eq:H6_general} therefore becomes time independent and reduces to
\begin{equation}
\boldsymbol H
=
\frac{\hbar}{2}
\begin{pmatrix}
0 & \Omega_P & 0 & 0 & 0 & 0 \\
\Omega_P & 0 & \Omega_C & 0 & 0 & 0 \\
0 & \Omega_C & 0 & \Omega_1 & 0 & \Omega_4 \\
0 & 0 & \Omega_1 & 0 & \Omega_2 & 0 \\
0 & 0 & 0 & \Omega_2 & 0 & \Omega_3 \\
0 & 0 & \Omega_4 & 0 & \Omega_3 & 0
\end{pmatrix}.
\label{eq:H6_res}
\end{equation}

To incorporate environmental interactions, the atomic ensemble is represented by the density matrix $\boldsymbol\rho(t)$ defined as

\begin{equation}
\boldsymbol\rho(t)
=
[\rho_{i,j}(t)]_{i,j=1}^{6}
\in
\mathbb{C}^{6\times6},
\label{eq:density}
\end{equation}
where $\rho_{i,j}(t)$ denotes the coherence between states $|i\rangle$ and $|j\rangle$, 
while the diagonal elements $\rho_{i,i}(t)$ represent the populations of the corresponding atomic levels. The density matrix satisfies the physical constraints
\(
\boldsymbol\rho=\boldsymbol\rho^\dagger\), \(
\operatorname{Tr}(\boldsymbol\rho)=1\), and \(\boldsymbol\rho\succeq0\), 
which ensure Hermiticity, unit trace, and positivity of the quantum state. 
The evolution of $\boldsymbol\rho(t)$ is governed by the Lindblad master equation, for weak Markovian system-environment interactions \cite{PRXQuantum.5.020202} which is
\begin{align}
\dot{\boldsymbol\rho}(&t)
=
-\frac{\mathbbm{i}}{\hbar}\big[\boldsymbol H,\boldsymbol\rho(t)\big]
\nonumber\\
&+
\!\! \! \! \sum_{(\ell,k)\in\mathcal D}
\! \! \!\gamma_{\ell,k}
\left(
\boldsymbol L_{\ell,k}\boldsymbol\rho(t)\boldsymbol L_{\ell,k}^\dagger
-\frac{1}{2}
\left\{
\boldsymbol L_{\ell,k}^\dagger \boldsymbol L_{\ell,k},
\boldsymbol\rho(t)
\right\}
\right),
\label{eq:Lindblad}
\end{align}


\noindent where $\boldsymbol L_{\ell,k} = |k\rangle\langle\ell|$ is the jump 
operator for the irreversible decay $|\ell\rangle \to |k\rangle$ at 
rate $\gamma_{\ell, k}$~\cite{PRXQuantum.5.020202}, and 

\begin{equation}
   \mathcal{D} = \{(n+1,\, n) : 1 \leq n \leq 5\} \cup \{(6,3)\} 
\end{equation}
contains the dominant decay pathways of the system. Equation~\eqref{eq:Lindblad} therefore captures both  coherent evolution induced by the Hamiltonian \eqref{eq:H6_res} and  dissipative dynamics arising from the interaction of the atomic ensemble with its environment.

For analytical and numerical convenience, \eqref{eq:Lindblad} can be column order vectorized in Liouville space as \cite{PRXQuantum.5.020202}

\begin{equation}
\operatorname{vec}\!\big(\dot{\boldsymbol\rho}(t)\big)
=
\boldsymbol{\mathcal L}\,
\operatorname{vec}\!\big(\boldsymbol\rho(t)\big),
\label{eq:Liouville_vec}
\end{equation}
where the Liouvillian superoperator, applicable with column-ordered density matrix vectorization, takes the form

\begin{align}
\boldsymbol{\mathcal L}
=&
-\frac{\mathbbm{i}}{\hbar}
\big(
\boldsymbol I\otimes \boldsymbol H
-
\boldsymbol H^{T}\otimes \boldsymbol I
\big)
+\sum_{(\ell, k)\in\mathcal D}
\gamma_{\ell , k}
\boldsymbol L_{\ell, k}^{*}\otimes \boldsymbol L_{\ell, k}
\nonumber\\
&
-\frac{1}{2}
\sum_{(\ell, k)\in\mathcal D}
\gamma_{\ell , k}
\big(
\boldsymbol I\otimes \boldsymbol L_{\ell, k}^{\dagger}\boldsymbol L_{\ell, k}
+
\boldsymbol L_{\ell, k}^{T}\boldsymbol L_{\ell, k}^{*}\otimes \boldsymbol I
\big).
\label{eq:Liouvillian}
\end{align}

The steady-state density matrix $\boldsymbol\rho_{\mathrm{ss}}$ is obtained from the null space of the Liouvillian, i.e., $\boldsymbol{\mathcal L}\,\operatorname{vec}\!(\boldsymbol\rho_{\mathrm{ss}})=0$, subject to the physical constraints
\(
\operatorname{Tr}(\boldsymbol\rho_{\mathrm{ss}})=1\) and \(\boldsymbol\rho_{\mathrm{ss}}=\boldsymbol\rho_{\mathrm{ss}}^\dagger\). 
Once the steady-state solution is obtained, the optical coherence $\rho_{2,1}$ can be extracted and used to characterize the probe response of the six-level H-RAQR.

\begin{Remark}[Model assumptions and practical broadening mechanisms]
\label{re:practical}

The analytical framework developed in this paper assumes resonant
excitation and neglects Doppler broadening, collision-induced
dephasing, transit-time broadening, laser linewidth, and field
inhomogeneity. These effects become important when quantitatively
predicting the transmission spectrum of practical room-temperature
vapor-cell receivers. Their inclusion requires velocity-dependent
density-matrix analysis and additional dephasing channels in the
Lindblad master equation. Consistent with existing communication-level
models of Rydberg receivers
\cite{cui2026rydbergatomicreceiversmultiband,Gong2025AnEB,gong2025rydbergatomicquantumreceivers},
these effects are omitted here to establish a baseline analytical 
atom-field interaction model and the corresponding communication 
framework for the proposed hybrid receiver, and are deferred to 
future work.
\end{Remark}

\subsection{Closed-Form Expression for the Probe Coherence}
\label{sec:analytical_rho12}

To characterize the optical response of the six-level H-RAQR, we focus on the steady-state coherence $\rho_{2,1}$ associated with the probe transition $|1\rangle \leftrightarrow |2\rangle$.  As the primary observable for optical probe transmission, $\rho_{2,1}$ directly determines the sensitivity of the receiver to external RF fields.
To obtain a tractable closed-form expression, we adopt the commonly used regime in which the decay rate from $|2\rangle$ is dominant, namely $\gamma_{2,1} \gg \gamma_{3,2},\,\gamma_{4,3},\,\gamma_{5,4},\,\gamma_{6,5},\,\gamma_{6,3}$, as reported in prior studies \cite{cui2026rydbergatomicreceiversmultiband,Jing2020}. Under this approximation, the dissipative dynamics are dominated by the channel $|2\rangle \to |1\rangle$, while the remaining higher-state decay processes are neglected in the analytical treatment. This yields a simplified but physically transparent model for the steady-state optical response.

Under resonant conditions in \eqref{eq:resonant_cond} and the dominant-decay approximation, the steady-state probe coherence is obtained as
\begin{equation}
\rho_{2,1} = 
\frac{-\mathbbm{i} \Omega_P \gamma_{2,1} \zeta^2}
{ \gamma_{2,1}^2 \zeta^2
+2 \Omega_P^4 \! \sum_{n=1}^{4} \Omega_n^2 
+ \! 2 \left[ \left(\Omega_2^2 \! + \! \Omega_3^2\right) \Omega_C^2 \! + \! \zeta^2 \right] \Omega_P^2 },
\label{eq:rho21_steady}
\end{equation}

\noindent where $\zeta = \Omega_1\Omega_3 - \Omega_2\Omega_4$ represents the interference between the cascaded excitation pathway
$\mathcal{T}_1\!\rightarrow\!\mathcal{T}_2\!\rightarrow\!\mathcal{T}_3$
and the non-adjacent pathway $\mathcal{T}_4$, thereby capturing the
characteristic hybrid coupling mechanism of the proposed H-RAQR. Here,
$\Omega_n$, $n\in\{1,2,3,4\}$, denote the effective RF Rabi frequencies
associated with the upper Rydberg transitions. To the best of our
knowledge, \eqref{eq:rho21_steady} is the first closed-form
steady-state probe-coherence expression reported for a hybrid PRS--CRS
Rydberg receiver architecture.

Equation~\eqref{eq:rho21_steady} is valid under the resonant and
dominant-decay assumptions adopted throughout this section, provided
the operating point lies outside the weak-coupling and
interference-balanced regimes identified in
Section~\ref{sec:validityandLimitation}. Within this validity region,
\eqref{eq:rho21_steady} accurately predicts the probe coherence, as
confirmed by the fidelity analysis in
Section~\ref{sec:F_operating} and the numerical validation in
Section~\ref{subsec:steady_state_validation}. The complete analytical
steady-state density matrix is provided in
Appendix~\ref{app:density-matrix}.

\subsection{Validity and Limitations of the Analytical Model}
\label{sec:validityandLimitation}

To assess the validity of the closed-form analytical expression derived in
Section~\ref{sec:analytical_rho12}, the steady-state probe coherence term
$\rho_{2,1}$ is compared with numerical solutions obtained from the Liouvillian superoperator in~\eqref{eq:Liouvillian}. 
The analytical model assumes resonant excitation
($\Delta_n=0$, $n\in\{P,C,1,2,3,4\}$) and retains only the dominant decay
channel $\gamma_{2,1}$, whereas the numerical simulations incorporate
experimentally realistic conditions including finite detunings ($\Delta_n \neq 0$, $n \in \{P,C,1,2,3,4\}$), and the full
set of decay channels
\(
\{\gamma_{i,j} : (i,j)\in\mathcal{D}\}.
\)
The upper-state decay rates are taken to be several orders of magnitude
smaller than $\gamma_{2,1}$, consistent with experimentally reported
parameters for RAQRs
\cite{cui2026rydbergatomicreceiversmultiband,Jing2020}.
This comparison provides a systematic evaluation of the robustness of the
analytical model under realistic experimental conditions.

The analysis reveals two parameter regimes where the simplified analytical
model exhibits noticeable deviations from the full numerical solution.

\begin{enumerate}

\item\textbf{Weak RF–Coupling Regime ($\Omega_2,\Omega_3\rightarrow0$).}
When RF coupling strengths $\Omega_2$ and $\Omega_3$ become very small,
the analytical model predicts $\rho_{2,1}\rightarrow0$, whereas the numerical
solution of the full master equation remains finite.
This discrepancy originates from the simplifying assumption that neglects
decay processes in the upper Rydberg manifold.
In the analytical model, the absence of upper-manifold decay channels allows population to accumulate unphysically in the upper Rydberg states when $\Omega_2$ and $\Omega_3$ are insufficient to couple them back to the lower levels, leaving population trapped with no dissipative 
pathway out. In contrast, the numerical model includes weak but finite decay processes
that redistribute population back toward the lower levels, thereby stabilizing
the steady state with a nonzero probe coherence ($\rho_{2,1}\neq0$).

\item\textbf{Interference-Balanced Regime ($\Omega_1\Omega_3 \approx
\Omega_2\Omega_4$).}
A second regime arises when the condition
\(
\Omega_1\Omega_3 \approx \Omega_2\Omega_4
\)
is satisfied. 
This relation corresponds to destructive interference between the sequential pathway
\(
|3\rangle \!\leftrightarrow\! |4\rangle
\!\leftrightarrow\! |5\rangle
\!\leftrightarrow\! |6\rangle
\)
and the direct transition
\(
|3\rangle \!\leftrightarrow\! |6\rangle .
\)
Under this interference condition the effective coupling between the Rydberg
manifold and the lower optical states vanishes, leading to
$\rho_{2,1}\rightarrow0$.
In this regime the analytical and numerical predictions for $\rho_{2,1}$ are in
agreement. 
However, numerical simulations reveal slow relaxation dynamics within the
Rydberg subspace
$\{|3\rangle,|4\rangle,|5\rangle,|6\rangle\}$ due to the formation of a
closed coherent loop.
Because dissipative processes in this manifold are weak, the system 
exhibits long-lived transient oscillations that eventually decay as 
$t\rightarrow\infty$, settling into a steady state in close agreement 
with the analytically derived result.

\end{enumerate}

These observations indicate that the closed-form analytical model accurately describes probe coherence over a broad range of RF coupling
strengths. 
Nevertheless, a fully general analytical treatment valid for arbitrary RF
couplings $\{\Omega_1,\Omega_2,\Omega_3,\Omega_4\}$ would require including additional relaxation channels associated with upper
Rydberg states.

\begin{Remark}[RF-loop interference invariant]
\label{re:zeta}
The quantity 
\(
\zeta = \Omega_1\Omega_3-\Omega_2\Omega_4
\)
naturally emerges in the analytical solution because it represents the
net coherent coupling imbalance around the RF loop
\(
|3\rangle \leftrightarrow |4\rangle \leftrightarrow |5\rangle \leftrightarrow |6\rangle \leftrightarrow |3\rangle .
\)
When $\zeta=0$, these pathways interfere destructively and the effective
coupling between the lower optical subsystem and the upper Rydberg
manifold vanishes. This condition therefore suppresses the probe
coherence $\rho_{2,1}$ and corresponds to the interference-balanced
regime discussed above.
\end{Remark}

While a more general solution can be obtained by retaining all parameters, the resulting symbolic expressions become algebraically cumbersome and provide limited analytical insight. The reduced analytical model therefore serves as a tractable approximation that captures the dominant system behavior.

The validity analysis above also indicates that the accuracy of the reduced analytical model represented by the closed-form probe coherence in \eqref{eq:rho21_steady} depends on the operating point in the Rabi-frequency space $\{\Omega_1,\Omega_2,\Omega_3,\Omega_4\}$. In practical receivers, strong LO fields establish a fixed operating bias for these transitions, while the received RF signals introduce only small perturbations around this point. Consequently, the analytical model needs to remain accurate only within a narrow neighborhood of the LO operating point. 
The operating regime in which the reduced analytical model accurately reproduces the steady-state dynamics of the full Lindblad system is identified next using a fidelity-based criterion.

\section{Fidelity-Based Identification of the Valid Operating Regime}
\label{sec:F_operating}

The analytical steady-state solution in \eqref{eq:rho21_steady} is derived under the approximation that the dominant decay channel $\gamma_{2,1}$ governs the dissipative dynamics while higher-order relaxation processes are neglected. As discussed in Section~\ref{sec:validityandLimitation}, this reduced model accurately describes the system dynamics only within certain regions of the RF Rabi-frequency space.

To identify the operating conditions under which the reduced analytical 
model remains accurate, we compare its predicted steady-state density 
matrix with the numerically obtained density matrix. Let
$\boldsymbol{\rho}_{\text{A}} \in \mathbb{C}^{6\times6}$ denote the 
steady-state density matrix obtained from the reduced analytical model, 
whose elements are given in Appendix~\ref{app:density-matrix}. Let
$\boldsymbol{\rho}_{\text{N}} \in \mathbb{C}^{6\times6}$ denote the 
numerically obtained density matrix, computed by integrating the 
Liouvillian superoperator time evolution in \eqref{eq:Liouville_vec}, retaining all decay channels and 
finite detunings. The solution is evaluated at $t = 10\,\mu$s, which 
is sufficient for convergence in all non-degenerate parameter regimes 
considered. The agreement between $\boldsymbol{\rho}_{\text{A}}$ and 
$\boldsymbol{\rho}_{\text{N}}$ is quantified using the quantum-state 
fidelity \cite{Jozsa1994}

\begin{equation}
\mathcal{F}(\boldsymbol{\rho}_{\text{N}},\boldsymbol{\rho}_{\text{A}})
=
\left[
\mathrm{Tr}
\left(
\sqrt{
\sqrt{\boldsymbol{\rho}_{\text{A}}}
\boldsymbol{\rho}_{\text{N}}
\sqrt{\boldsymbol{\rho}_{\text{A}}}
}
\right)
\right]^2 .
\label{eq:fidelity}
\end{equation}

As visualized in Fig.~\ref{fig:fidelityFull}, which is presented in Section~\ref{subsec:steady_state_validation}, the fidelity landscape across the four-dimensional Rabi-frequency space is predominantly near unity (white in~Fig.~\ref{fig:fidelityFull}), indicating that the reduced analytical model agrees closely with the full Lindblad solution over most of the accessible parameter space. Reduced fidelity (darker regions) appears only along the diagonal strips satisfying $\Omega_1 \Omega_3 \approx \Omega_2 \Omega_4$  and near the weak-coupling \(\Omega_2,\Omega_3 \to 0\) boundaries, consistent with the two breakdown regimes identified in Section~\ref{sec:validityandLimitation}.

\subsection{LO Operating Point and Small-Signal Excursions}

In practical receiver operation, the RF transitions are driven by strong local-oscillator (LO) fields that establish a fixed operating point in the Rabi-frequency space. The received RF signals introduce only small perturbations around this operating point. Accordingly, the effective Rabi frequencies can be written as
\begin{align}
&\Omega_n(t)
=
\Omega^{\star}_{\mathrm{LO},n}
+
\tilde{\Omega}_{\mathrm{RF},n}(t),
\nonumber\\
&\qquad
|\tilde{\Omega}_{\mathrm{RF},n}| \ll \Omega^{\star}_{\mathrm{LO},n},
\quad n=1,\dots,4,
\label{eq:rabi_operating}
\end{align}
where $\Omega^{\star}_{\mathrm{LO},n}$ denotes the LO-induced Rabi frequency of 
the $n$-th RF transition, and $\tilde{\Omega}_{\mathrm{RF},n}(t) = 
\frac{\mu_n}{\hbar}A_{\mathrm{RF},n}\cos(\delta\omega_n t + \delta\phi_n)$ 
is the small perturbation induced by the received RF signal as defined 
in~\eqref{eq:rabi_heterodyne_second}. 

Since the steady-state density matrix depends only on the instantaneous Rabi amplitudes, the  time dependence is omitted for notational simplicity. We therefore define the Rabi-frequency vector and LO operating point, respectively, as 
\begin{align}
\boldsymbol{\Omega}
&=
(\Omega_1,\Omega_2,\Omega_3,\Omega_4)^{T}
\in \mathbb{R}^{4},
\nonumber\\
\boldsymbol{\Omega}^{\star}_{\mathrm{LO}}
&=
(\Omega^{\star}_{\mathrm{LO},1},
\Omega^{\star}_{\mathrm{LO},2},
\Omega^{\star}_{\mathrm{LO},3},
\Omega^{\star}_{\mathrm{LO},4})^{T} \in \mathbb{R}^{4}.
\end{align}
The received RF signals induce small excursions around this operating point. We express this compactly as
\(\boldsymbol{\Omega}
=
\boldsymbol{\Omega}^{\star}_{\mathrm{LO}}
+
\boldsymbol{\epsilon}\),
where $\boldsymbol{\epsilon}$ represents a small perturbation vector. The admissible perturbation region is modeled as
\begin{equation}
\mathcal{R}\hspace{-0.5mm}
=\hspace{-0.5mm}
\left\{
\boldsymbol{\Omega}\in\mathbb{R}^{4} :
|\Omega_n-\Omega^{\star}_{\mathrm{LO},n}|
\le \Delta\Omega_n,
\;\hspace{-0.5mm} n=1,\dots,4
\right\},
\label{eq:region}
\end{equation}
where $\Delta\Omega_n$ denotes the maximum RF-induced excursion in the $n$-th Rabi frequency.

\subsection{Operating-Point Identification}
\label{subsec:operating_point}

The objective is to identify a LO operating point at
which the reduced analytical model accurately reproduces the steady-state
response of the full Lindblad system over the small-signal operating
region. Since the received RF signals act as small perturbations around
the LO operating point, only the LO Rabi frequencies are optimized,
while the received RF Rabi frequencies are treated as perturbations
within the neighborhood $\mathcal{R}$.

Formally, the operating point
$\boldsymbol{\Omega}^{\star}_{\mathrm{LO}}$ is obtained by maximizing
the average quantum-state fidelity over $\mathcal{R}$,
\begin{subequations}
\begin{align}
\max_{\boldsymbol{\Omega}_{\mathrm{LO}}}
\quad
&
\overline{\mathcal{F}}
=
\frac{1}{|\mathcal{R}|}
\int_{\mathcal{R}}
\mathcal{F}
\!\left(
\boldsymbol{\rho}_{\mathrm{N}},
\boldsymbol{\rho}_{\mathrm{A}}
\right)
d\boldsymbol{\Omega},
\\
\text{s.t.}\quad
&
\sum_{n=1}^{4}
\Omega_{\mathrm{LO},n}
\le
\Sigma\Omega_{\max},
\end{align}
\label{eq:opt_problem}
\end{subequations}
where $|\mathcal{R}|$ denotes the hypervolume of the perturbation region
defined in \eqref{eq:region}, and
$\Sigma\Omega_{\max}$ is the maximum available LO driving strength. 
Problem~\eqref{eq:opt_problem} is solved using the exhaustive search
procedure summarized in Algorithm~\ref{alg:LO_selection}. Since all
feasible operating points are examined, the identified solution is
independent of initialization.

\begin{algorithm}[t]
\caption{LO Operating-Point Identification}
\label{alg:LO_selection}
\begin{algorithmic}[1]
\REQUIRE Candidate LO search set $\mathcal{S}_{\Omega}$, perturbation region $\mathcal{R}$, maximum LO driving strength $\Sigma\Omega_{\max}$
\ENSURE Selected operating point $\boldsymbol{\Omega}^{\star}_{\mathrm{LO}}$

\STATE Initialize $\overline{\mathcal{F}}_{\max}\leftarrow0$ and
$\boldsymbol{\Omega}^{\star}_{\mathrm{LO}}\leftarrow\emptyset$.

\FOR{each candidate
$\boldsymbol{\Omega}_{\mathrm{LO}}\in\mathcal{S}_{\Omega}$}

    \IF{$\sum_{n=1}^{4}\Omega_{\mathrm{LO},n}\leq\Sigma\Omega_{\max}$}

    \STATE Compute $\boldsymbol{\rho}_{\mathrm{A}}(\boldsymbol{\Omega})$ via   Appendix~\ref{app:density-matrix} for each $\boldsymbol{\Omega} \in \mathcal{R}$.

\STATE Compute $\boldsymbol{\rho}_{\mathrm{N}}(\boldsymbol{\Omega})$ 
by solving \eqref{eq:Lindblad} for each $\boldsymbol{\Omega} \in \mathcal{R}$.

    

        \STATE Evaluate the fidelity
        $\mathcal{F}(\boldsymbol{\rho}_{\mathrm N},
        \boldsymbol{\rho}_{\mathrm A})$
        at every sample in $\mathcal{R}$.

        \STATE Compute the average fidelity
        $\overline{\mathcal{F}}$.

        \IF{$\overline{\mathcal{F}}>\overline{\mathcal{F}}_{\max}$}

            \STATE Update
            $\overline{\mathcal{F}}_{\max}
            \leftarrow
            \overline{\mathcal{F}}$.

            \STATE Update
            $\boldsymbol{\Omega}^{\star}_{\mathrm{LO}}
            \leftarrow
            \boldsymbol{\Omega}_{\mathrm{LO}}$.

        \ENDIF

    \ENDIF

\ENDFOR

\STATE If multiple operating points achieve the same
$\overline{\mathcal{F}}_{\max}$, select the one requiring the smallest
total LO driving strength.

\RETURN $\boldsymbol{\Omega}^{\star}_{\mathrm{LO}}$.

\end{algorithmic}
\end{algorithm}

The exhaustive search guarantees reproducibility and eliminates any
dependence on initialization. In practice, the optimum is found within
a connected high-fidelity plateau rather than at an isolated point,
indicating the existence of multiple operating points with nearly
identical fidelity. Selecting the minimum-power solution within this
plateau provides additional robustness while ensuring that RF-induced
perturbations remain inside the validity region of the analytical model
and away from the weak-coupling and interference-balanced breakdown
regimes identified in
Section~\ref{sec:validityandLimitation}. The selected operating point is
reported in Section~\ref{subsec:steady_state_validation}, while the
corresponding high-fidelity region is illustrated in
Appendix~\ref{app:fidelity}.

\section{Communication Signal Model and Performance Analysis}
\label{sec:communication_model}


This section derives the photodetector signal, the baseband channel 
representation, and the achievable communication performance of the 
proposed H-RAQR.

\subsection{Probe Transmission and Photodetection Model}
\label{subsec:probe_transmission}

The information carried by the incident RF fields is extracted optically by monitoring the transmitted probe laser intensity after propagation through the vapor cell. Following the power-attenuation framework commonly adopted in Rydberg atomic receivers~\cite{cui2026rydbergatomicreceiversmultiband,10972179,Gong2025AnEB}, the RF-induced modification of the atomic susceptibility is mapped onto variations of the probe transmission through the optical coherence $\rho_{2,1}$.

Under the adiabatic optical approximation, the probe field at the output of the vapor cell, can be expressed as~\cite{Gong2025AnEB,Meyer2021}

\begin{equation}
\mathcal{E}_{out}(t)
= \sqrt{P_0}\,
e^{\Xi_0\,\mathrm{Im}\{\rho_{2,1}\}}
\cos\!\big(\omega_P t-\Xi_0\,\mathrm{Re}\{\rho_{2,1}\}\big),
\label{eq:probe_field}
\end{equation}

\noindent where $P_0$ is input probe power and $\Xi_0 =
\frac{2\pi d N_0 \mu_P^2}{\varepsilon_0 \hbar \lambda_P \Omega_P}$ denotes the atomic susceptibility constant. Here $N_0$ is the atomic number density, $\mu_P$ is the dipole moment of the probe transition, $\lambda_P$ is the probe wavelength, and $d$ is the vapor-cell length.

The transmitted probe beam is detected using a photodetector whose output is proportional to the time-averaged optical intensity~\cite{SalehTeich2007,DiStefano2018}. Since the optical frequency $\omega_P$ is several orders of magnitude larger than the RF modulation frequencies, the detector measures the envelope power, i.e., $y(t)\propto\langle |\mathcal{E}_{out}(t)|^2\rangle$. 
Substituting \eqref{eq:probe_field} yields the nonlinear electrical response
\begin{equation}
y(\boldsymbol{\Omega})
=
R\,\frac{P_0}{2}
\exp\!\big[2\Xi_0\,\mathrm{Im}\{\rho_{2,1}\}\big],
\label{eq:photodetector_output}
\end{equation}
where $R$ denotes the photodetector responsivity and $\boldsymbol{\Omega}=(\Omega_1,\Omega_2,\Omega_3,\Omega_4)^T$ is the vector of RF Rabi frequencies. The dependence on $\boldsymbol{\Omega}$ arises through $\rho_{2,1}$.

To obtain a tractable communication model, the detector response is linearized around the LO operating point $\boldsymbol{\Omega}^{\star}_{\mathrm{LO}}$ determined in Section~\ref{subsec:operating_point}. Applying a first-order Taylor expansion of \eqref{eq:photodetector_output} gives
\begin{equation}
y(\boldsymbol{\Omega})
\approx
y_{\mathrm{LO}}
+
\sum_{n=1}^{4}
\left.
\frac{\partial y(\boldsymbol{\Omega})}{\partial \Omega_n}
\right|_{\boldsymbol{\Omega}^{\star}_{\mathrm{LO}}}
(\Omega_n-\Omega^{\star}_{\mathrm{LO},n}),
\label{eq:taylor_expansion}
\end{equation}
where $y_{\mathrm{LO}}=y(\boldsymbol{\Omega}^{\star}_{\mathrm{LO}})$ is the steady photodetector output in the absence of received RF signals.

Since the received RF fields are much weaker than the LO drives ($\Omega_{\mathrm{RF},n}\ll\Omega^{\star}_{\mathrm{LO},n}$),  substituting the instantaneous Rabi frequency established 
in \eqref{eq:rabi_operating} into 
\eqref{eq:taylor_expansion} yields the linearized photodetector output
\begin{equation}
y(t)\approx
y_{\mathrm{LO}}
+
\sum_{n=1}^{4}
\mathcal{G}_n A_{\mathrm{RF},n}
\cos(\delta\omega_n t+\delta\phi_n)
+
w(t),
\label{eq:linear_PD_output}
\end{equation}
where $w(t)$ denotes additive detector noise and
\begin{equation}
    \mathcal{G}_n=
\left.
\frac{\mu_n}{\hbar}
\frac{\partial y(\boldsymbol{\Omega})}{\partial \Omega_n}
\right|_{\boldsymbol{\Omega}^{\star}_{\mathrm{LO}}}
\end{equation}
is the RF-to-electrical gain coefficient of the atomic receiver.

\subsection{Equivalent Baseband Communication Model}
\label{subsec:baseband_model}

\begin{figure}
\includegraphics[width=1\linewidth]{ 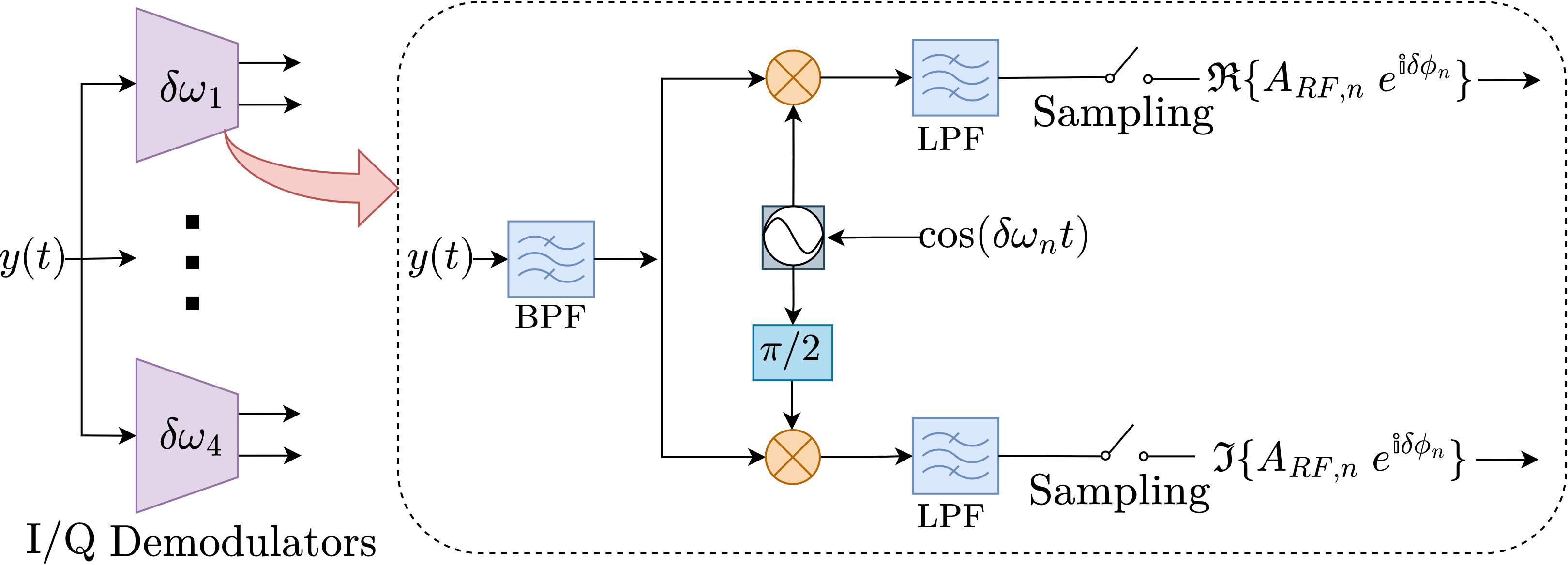}
\caption{Block diagram illustrating multi-channel in-phase and quadrature (I/Q) demodulation of the photodetector output at the intermediate frequencies $\delta\omega_n$.\label{Demodulation}}
\end{figure}

The photodetector output $y(t)$ in \eqref{eq:linear_PD_output} is processed to extract the complex baseband components of each RF channel. As illustrated in Fig.~\ref{Demodulation}, the PD output is first split into multiple branches, each of which is passed through a bandpass filter (BPF) centered at $\delta\omega_n$. The filtered signals are subsequently coherently down-converted using in-phase and quadrature reference tones, followed by low-pass filtering (LPF) ~\cite{cui2026rydbergatomicreceiversmultiband,Gong2025AnEB}. This standard I/Q demodulation extracts the complex baseband component of the photodetector output, which can be expressed as $y_n = \mathcal{G}_n s_n e^{\mathbbm{i}\delta\phi_n} + w_n$, where $\delta\phi_n$ denotes the phase offset between the received RF signal and the corresponding local oscillator, $s_n$ is the complex envelope of the received RF field, and $w_n$ represents the equivalent complex baseband noise.

For the $n$th channel, the received RF complex envelope is modeled as $s_n = \sqrt{P_{T,n}}\, h_n x_n$, where $P_{T,n}$ denotes the transmitted RF power, $h_n$ is the complex wireless channel coefficient, and $x_n$ is the transmitted symbol satisfying $\mathbb{E}[|x_n|^2]=1$. Absorbing the phase offset into the effective channel definition $\tilde h_n = h_n e^{\mathbbm{i}\delta\phi_n}$ yields the equivalent baseband signal model
\begin{equation}
y_n
=
\mathcal{G}_n \sqrt{P_{T,n}}\, \tilde h_n\, x_n
+
w_n,
\qquad n=1,\dots,4 ,
\label{eq:basebandSignalmodel}
\end{equation}
which corresponds to a standard complex baseband channel where $\mathcal{G}_n$ represents the RF-to-electrical conversion gain and $\tilde h_n$ denotes the effective wireless channel.

The performance of the receiver is governed by the signal-to-noise ratio (SNR), which depends on the dominant noise processes. Following the noise modeling framework commonly adopted for Rydberg atomic receivers~\cite{Gong2025AnEB,cui2026rydbergatomicreceiversmultiband,Tu2024}, the equivalent baseband noise in \eqref{eq:basebandSignalmodel} is modeled as a complex Gaussian random variable $w_n \sim \mathcal{CN}(0,\sigma_n^2)$, assumed to remain 
statistically stationary over the observation interval and 
independent across channels. The noise variance $\sigma_n^2 = \sigma_{I,n}^2 + \sigma_{E,n}^2$ represents the aggregate contribution of intrinsic optical noise and environmental RF noise, where $\sigma_{I,n}^2$ corresponds to photon shot noise at the photodetector and $\sigma_{E,n}^2$ accounts for RF noise induced by ambient black-body radiation.

The intrinsic noise component scales with the detected probe power and is given by $\sigma_{I,n}^2 = y_{\mathrm{LO}} B_n \hbar \omega_P$, where $y_{\mathrm{LO}}$ denotes the steady photodetector output in the absence of RF signals, $B_n$ is the baseband bandwidth of the $n$th channel, and $\omega_P$ is the probe optical frequency~\cite{Tu2024}. The extrinsic noise arises from black-body radiation coupling to the Rydberg transitions ~\cite{Tu2024,cui2026rydbergatomicreceiversmultiband} and is modeled as $\sigma_{E,n}^2 = \mathcal{G}_n^{2} B_n S_{\mathrm{BB}}(\omega_n,T_a)$, where $\omega_n$ is the RF transition frequency, $T_a$ denotes the ambient 
temperature.  The corresponding black-body spectral density is
\begin{equation}
S_{\mathrm{BB}}(\omega_n,T_a)
=
\frac{4\hbar \omega_n^{3}}{\varepsilon_0 c^{3}}
\frac{e^{\hbar\omega_n/(k_B T_a)} + 1}
     {e^{\hbar\omega_n/(k_B T_a)} - 1},
\label{eq:bbr_spectral_density}
\end{equation}
where $k_B$ is the Boltzmann constant, $\varepsilon_0$ is the free-space 
permittivity, and $c$ is the speed of light~\cite{ScullyZubairy1997}. 
Using the baseband signal model in \eqref{eq:basebandSignalmodel} and assuming unit-power modulation $\mathbb{E}[|x_n|^2]=1$, the SNR of the $n$-th RF channel is given by
\begin{equation}
\mathrm{SNR}_n
=
\mathcal{G}_n^2 P_{T,n} |\tilde h_n|^2/\sigma_n^2,
\label{eq:snr}
\end{equation}
which forms the basis for the achievable rate analysis.

\subsection{Achievable Sum Rate}
\label{subsec:rate_analysis}

The hybrid six-level architecture supports multiple simultaneous RF channels. To quantify the aggregate communication capability of the receiver, we adopt the achievable sum-rate metric, which captures the total information throughput across all active RF transitions. This metric is particularly suitable for comparing different atomic receiver architectures (e.g., CRS, PRS, and hybrid systems) because it jointly reflects both the achievable SNR and the number of simultaneously supported channels. The instantaneous sum rate is therefore given by
\begin{equation}
R_{\mathrm{sum}}
=
\sum_{n=1}^{N_c}
B_n \log_2\!\left(1+\mathrm{SNR}_n\right),
\label{eq:sumrate}
\end{equation}
where $N_c$ denotes the number of active RF channels and $B_n$ is  bandwidth of the $n$th channel. 
Assuming independent Rayleigh fading channels $h_n \sim \mathcal{CN}(0,\beta_n)$, the ergodic sum rate of the receiver is obtained by averaging \eqref{eq:sumrate} over channel statistics
\begin{align}
\bar R_{\mathrm{sum}}
&=
\sum_{n=1}^{N_c}
B_n
\mathbb{E}_{|h_n|^2}
\!\left[
\log_2\!\left(
1+\frac{\mathcal{G}_n^2 P_{T,n}|h_n|^2}{\sigma_n^2}
\right)
\right] \notag \\
&=
\sum_{n=1}^{N_c}
\frac{B_n}{\ln 2}\,
e^{1/\bar\Gamma_n}
E_1\!\left(\frac{1}{\bar\Gamma_n}\right),
\label{eq:ergodic_sum_rate_closed}
\end{align}
where $\bar\Gamma_n = \mathcal{G}_n^2 P_{T,n} \beta_n / \sigma_n^2$ denotes 
the average SNR of $n$th channel, $E_1\!\left(\cdot\right)$ is the exponential integral function, and the closed-form result follows from standard ergodic capacity identity for Rayleigh fading 
channels~\cite{Goldsmith2005}.

\subsection{Resource-Efficiency Metric}
\label{subsec:resource_efficiency}

The ergodic sum rate in \eqref{eq:ergodic_sum_rate_closed}
characterizes the aggregate communication throughput of a receiver
architecture, but does not account for the resources required to realize
its RF-band coverage. Since the PRS and CRS schemes each access only a subset of the hybrid manifold, the H-RAQR's full four-band coverage can only be matched by jointly deploying both schemes, two independent vapor-cell front-ends, each with its own probe and coupling laser pair.

Specifically, the achievable sum rate is normalized by the total optical
resources consumed, comprising the number of vapor-cell front-ends
$N_{\mathrm{FE}}$ and the optical power
$P_{\mathrm{laser}}=P_P+P_C$ supplied to the probe and coupling lasers
of each front-end. The optical powers are recovered from the field
amplitudes in \eqref{eq:optical_rabi} as
\(
P_k=\frac{1}{2}\varepsilon_0 c\,\mathcal{A}\,A_k^2,
\;
k\in\{P,C\},
\)
where $\mathcal{A}$ denotes the beam cross-sectional area. The resulting
resource efficiency is defined as
\begin{equation}
\eta
=
\frac{\bar R_{\mathrm{sum}}}
{N_{\mathrm{FE}}P_{\mathrm{laser}}}
\quad \left[{\mathrm{bps}}/{\mathrm{W}}\right],
\label{eq:resource_efficiency}
\end{equation}
which measures the achievable sum rate per unit optical
receiver resource. This metric is used in
Section~\ref{subsec:resource_efficiency_results} to compare the proposed
H-RAQR with conventional PRS and CRS architectures under equivalent
multi-band reception capability.


\section{Numerical Results and Discussion}
\label{sec:results}

Numerical results are obtained using the full Lindblad master-equation
model of the proposed six-level H-RAQR based on a
$^{133}\mathrm{Cs}$ vapor-cell receiver. Unless otherwise specified,
all simulation parameters are adopted from experimentally reported
Rydberg receiver implementations and established theoretical
models~\cite{cui2026rydbergatomicreceiversmultiband,Gong2025AnEB,Jing2020}.


\subsection{Simulation Setup}

The atoms are confined in a room-temperature vapor cell of length $d = 2~\mathrm{cm}$ with atomic number density $N_0 = 4.89 \times 10^{10}~\mathrm{cm^{-3}}$.
The probe laser, operating at wavelength $\lambda_P = 852.35~\mathrm{nm}$, drives the transition $|1\rangle = 6S_{1/2} \rightarrow |2\rangle = 6P_{3/2}$ with Rabi frequency $\Omega_P = 2\pi \times 5.7~\mathrm{MHz}$. A coupling laser at $\lambda_C \approx 509.14~\mathrm{nm}$ excites the transition $|2\rangle \rightarrow |3\rangle$ with $\Omega_C = 2\pi \times 0.97~\mathrm{MHz}$, preparing the initial Rydberg state $|3\rangle = 60D_{5/2}$. 
The upper Rydberg manifold consists of the states $|4\rangle = 62P_{3/2}$, $|5\rangle = 61D_{5/2}$, and $|6\rangle = 60F_{7/2}$. This configuration enables both the cascaded RF pathway $|3\rangle \rightarrow |4\rangle \rightarrow |5\rangle \rightarrow |6\rangle$ and the direct coupling $|3\rangle \rightarrow |6\rangle$, forming the hybrid interaction structure proposed in this work.

The RF transition frequencies and dipole moments are computed using  ARC (Alkali Rydberg Calculator) package~\cite{Sibalic2017ARC}. The resulting RF transitions, corresponding communication bands, and dipole moments are summarized in Table~\ref{tab:rf}, where $e = 1.602 \times 10^{-19}~\mathrm{C}$ denotes the unit charge and $a_0 = 5.29 \times 10^{-11}~\mathrm{m}$ is the Bohr radius. We note that the considered application bands correspond to
narrowband communication links, consistent with the instantaneous
bandwidth limitations of EIT-based
RAQRs~\cite{PhysRevApplied.19.014025,11124471}. The detuning values in Table~\ref{tab:rf} correspond to a vanishing closed-loop RF detuning ($\delta=0$). Since the upper Rydberg decay rates are several orders of magnitude smaller than the optical decay rate $\gamma_{2 , 1}$, the relaxation parameters are chosen as
$\gamma_{2, 1}=2\pi\times5.2~\mathrm{MHz}$,
$\gamma_{3, 2}=2\pi\times3.9~\mathrm{kHz}$,
$\gamma_{4,3}=2\pi\times1.7~\mathrm{kHz}$,
$\gamma_{5, 4}=2\pi\times1.6~\mathrm{kHz}$,
$\gamma_{6, 5}=2\pi\times1.5~\mathrm{kHz}$,
and $\gamma_{6,3}=2\pi\times1.5~\mathrm{kHz}$.
For notational compactness, all Rabi frequencies in the remainder 
of this section are given in normalized units of $2\pi\times\mathrm{MHz}$, 
so that, e.g., $\Omega = 2\pi\times7~\mathrm{MHz}$ is written 
as $\Omega = 7$.


\begin{table}[t]
\centering
\caption{Rydberg RF transitions with frequencies,
dipole moments $\mu_n$ [$ea_0$], and detunings $\Delta_n$ 
[$2\pi\times\mathrm{kHz}$].}
\label{tab:rf}
\begin{tabular}{ccccc}
\hline
\textbf{Transition} & \textbf{Frequency}\textbf{($\omega_n/2\pi$)} & \textbf{$\mu_n$} \textbf{[$ea_0$]} & \textbf{$\Delta_n$} \textbf{[$2\pi$kHz]}\\
 &  & &  \\
\hline
$|3\rangle\rightarrow|4\rangle$  & $30.615~\mathrm{GHz}$ & $2329.67$ & $1$ \\
$|4\rangle\rightarrow|5\rangle$ & $3.054~\mathrm{GHz}$  & $7886.52$ & $1$ \\
$|5\rangle\rightarrow|6\rangle$ & $45.342~\mathrm{GHz}$ & $711.764$ & $2$ \\
$|3\rangle\rightarrow|6\rangle$  & $79.01~\mathrm{GHz}$  & $250.939$ & $4$ \\
\hline
\end{tabular}
\end{table}

\subsection{Validation of the Analytical Steady-State Model}
\label{subsec:steady_state_validation}

To validate the analytical steady-state solution derived in 
Sec.~\ref{sec:analytical_rho12}, we compare the analytical density matrix 
with the numerical solution obtained from the full Lindblad master equation. 
The numerical density matrix is extracted from the time evolution at 
$t=10~\mu\mathrm{s}$, which ensures convergence to steady state for all 
parameter regimes considered.

\begin{figure}[t]
    \centering
    \includegraphics[width=1\linewidth]{ 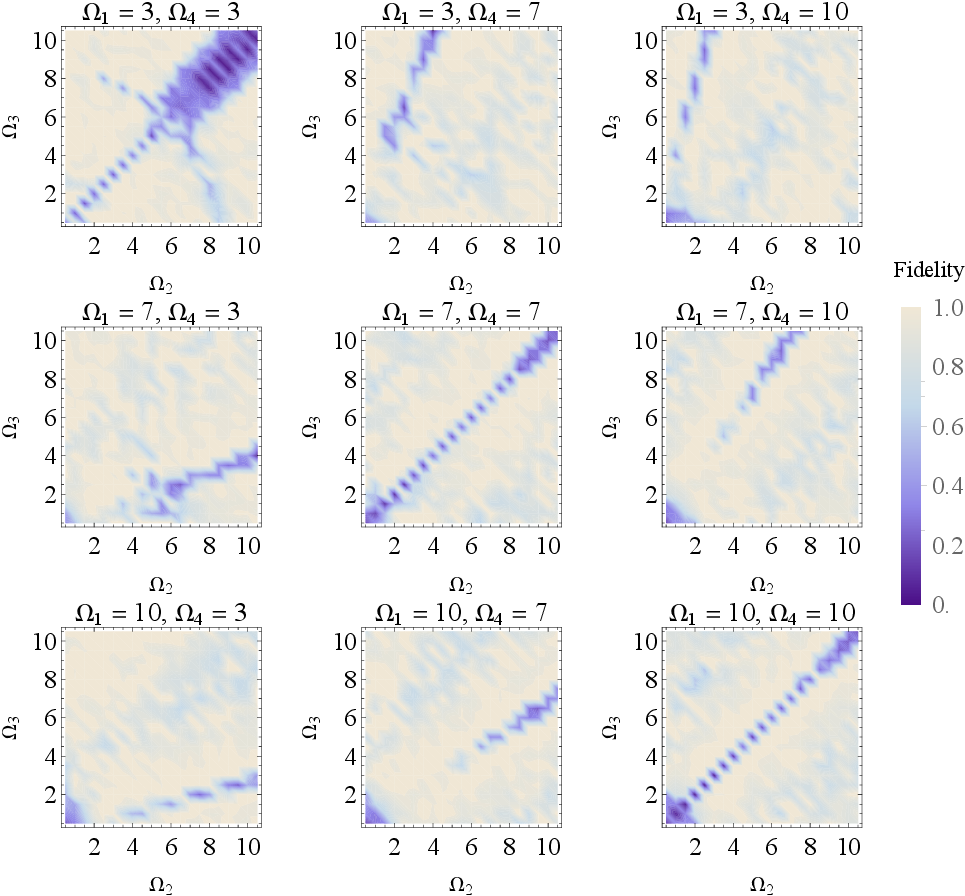}
    \caption{Fidelity landscape between the $\boldsymbol{\rho}_{\text{A}}$ and 
$\boldsymbol{\rho}_{\text{N}}$. Each subplot shows a two-dimensional slice of the 
four-dimensional Rabi-frequency space, where $\Omega_2$ and $\Omega_3$ are 
varied while $(\Omega_1,\Omega_4)$ are fixed to the values indicated above 
each panel. All Rabi frequencies are in units of $2\pi\times\mathrm{MHz}$.}
    \label{fig:fidelityFull}
\end{figure}

\subsubsection{Global Fidelity Landscape}

The agreement between the analytical and numerical solutions is quantified 
using the quantum-state fidelity $\mathcal{F}$ defined in 
Sec.~\ref{subsec:operating_point}. Since the fidelity depends on four RF 
Rabi frequencies $(\Omega_1,\Omega_2,\Omega_3,\Omega_4)$, direct visualization 
of the four-dimensional parameter space is not possible. Therefore, the 
fidelity landscape is examined through two-dimensional slices of the 
Rabi-frequency space. 
Fig.~\ref{fig:fidelityFull} presents a $3\times3$ grid of fidelity heatmaps. 
In each subplot, $\Omega_2$ and $\Omega_3$ are varied over the range 
$0\leq\Omega_2,\Omega_3\leq10$, while $(\Omega_1,\Omega_4)$ are fixed to the nine representative 
pairs $(3,3)$, $(3,7)$, $(3,10)$, $(7,3)$, $(7,7)$, $(7,10)$, 
$(10,3)$, $(10,7)$, and $(10,10)$.
Regions of reduced fidelity appear primarily in two regimes. The first occurs 
in the weak-coupling limit where the RF Rabi frequencies approach zero, making 
the neglected upper-manifold relaxation channels non-negligible. The second 
occurs near the interference-balanced condition 
$\Omega_1\Omega_3 \approx \Omega_2\Omega_4$, where coherent population trapping 
within the Rydberg manifold slows numerical convergence toward the steady 
state. Outside these regimes, the fidelity remains uniformly close to unity, 
indicating that the analytical steady-state model accurately reproduces the 
full Lindblad dynamics across most of the accessible parameter space.

\subsubsection{Operating-Point Identification}
\label{subsubsec:op_point}

Following Algorithm~\ref{alg:LO_selection}, an exhaustive grid search is
performed over the admissible LO Rabi-frequency space
$\boldsymbol{\Omega}_{\mathrm{LO}}\in[0,10]^4$ with a uniform step size
of $0.5$, subject to the power constraint
$\Sigma\Omega_{\max}=20$. For each feasible candidate, the average
fidelity in \eqref{eq:opt_problem} is evaluated over the perturbation
region $\mathcal{R}$ using
$\Delta\Omega_n=2\pi\times2~\mathrm{kHz}$, reflecting the small-signal
condition $\Omega_{\mathrm{RF}}\ll\Omega_{\mathrm{LO}}$. The resulting
operating point is
\begin{equation}
\boldsymbol{\Omega}^{\star}_{\mathrm{LO}}
=
\{2,\;7,\;1,\;6\},
\label{OptimumRabiLO}
\end{equation}
corresponding to
$\{\Omega_1,\Omega_2,\Omega_3,\Omega_4\}$, with
$\sum_{n}\Omega^{\star}_{\mathrm{LO},n}=16$, satisfying the prescribed
power budget. The identified operating point lies within a broad
high-fidelity plateau, achieving an average fidelity of
$\overline{\mathcal{F}}=0.999998$. This confirms that the analytical
steady-state model accurately reproduces the full Lindblad solution
throughout the local communication operating region. The 
fidelity landscape is shown in
Appendix~\ref{app:fidelity}.






\vspace{1mm}

\subsubsection{Population Dynamics and Steady-State Convergence}

Finally, we verify that the analytical steady-state solution also captures 
 full time-dependent quantum dynamics of the six-level system. The analytical model assumes resonant operation and retains only the dominant decay channel $\gamma_{2,1}$. Despite these simplifications, the resulting 
steady-state solution remains in good agreement with full numerical 
steady state, which accounts for all weak upper-state decay channels 
$\{\gamma_{i , j}\}$ $(\ll \gamma_{2, 1})$ and finite detunings.

\begin{figure}[t]

\centering
\includegraphics[width=0.45 \textwidth]{ 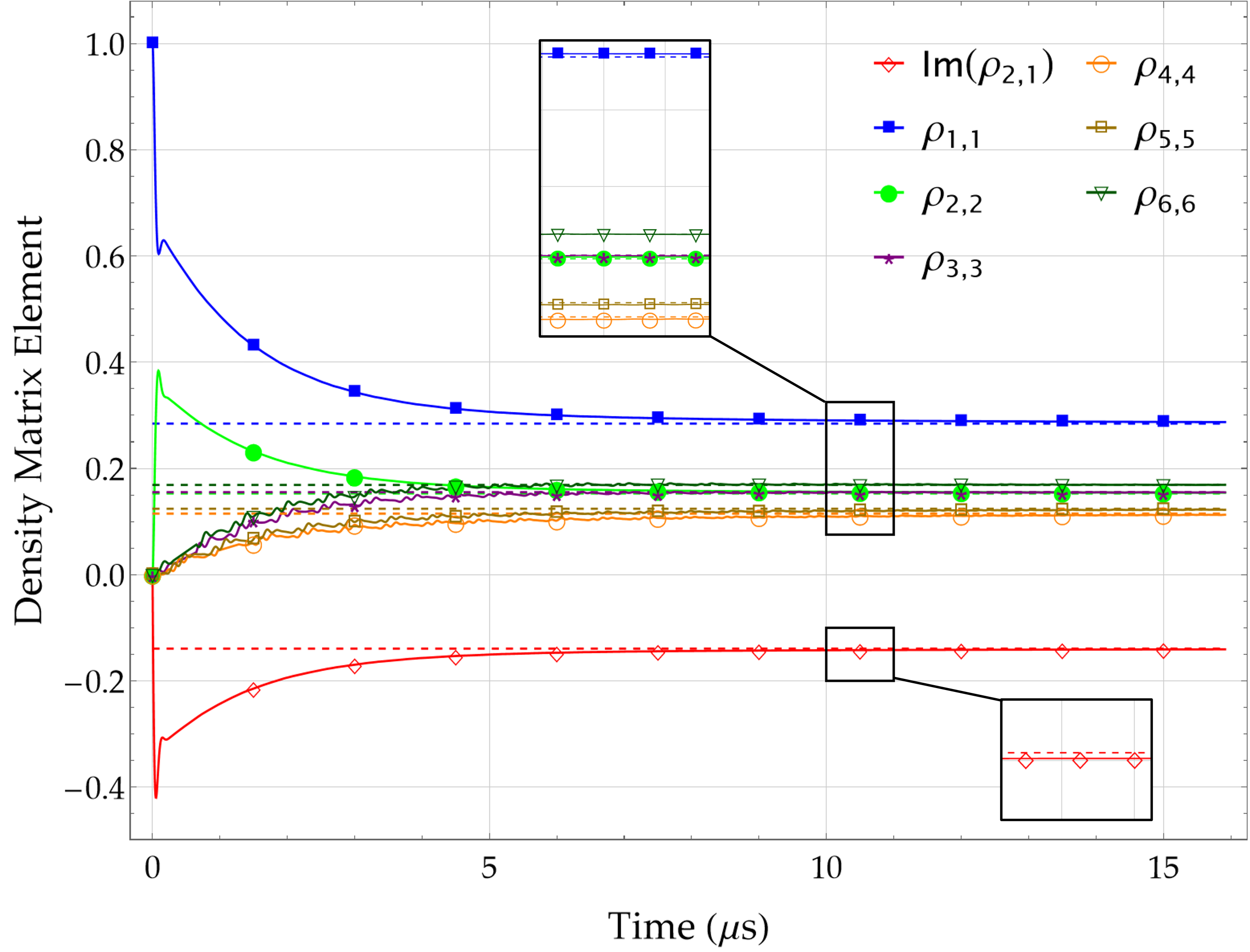}
\caption{Time evolution of the $\rho_{1,1}(t),\ldots,\rho_{6,6}(t)$ and the probe coherence
$\mathrm{Im}(\rho_{2,1}(t))$ at the optimized LO operating point
$\boldsymbol{\Omega}^{\star}_{\mathrm{LO}}$.
Solid curves show the numerical evolution.
The horizontal dashed lines represent the analytical steady-state
values predicted by \eqref{eq:rho21_steady}.
The convergence of numerical trajectories to these steady-state
levels confirms the validity of the analytical model.}
\label{fig:populations}
\end{figure}

Fig.~\ref{fig:populations} shows the numerically computed population 
evolution $\rho_{1,1}(t),\ldots,\rho_{6,6}(t)$ and probe coherence 
$\mathrm{Im}(\rho_{2,1}(t))$ obtained by solving the Lindblad superoperator in \eqref{eq:Liouvillian}. The curves represent the numerical time evolution, while the dashed horizontal lines indicate the 
analytical steady-state values predicted by \eqref{eq:rho21_steady}.
The simulation is performed at the optimized operating point 
$\boldsymbol{\Omega}^{\star}_{\mathrm{LO}}$ in \eqref{OptimumRabiLO}. As the system evolves, all 
population trajectories converge to the analytical steady-state values 
within approximately $10~\mu\mathrm{s}$. The close agreement between the 
numerical dynamics and the analytical predictions confirms that the reduced 
steady-state model accurately captures the dominant physical behavior of the 
six-level hybrid RAQR despite the presence of additional weak decay channels.


\begin{figure}[t]
\centering
\subfloat[\centering]{\includegraphics[width=0.241 \textwidth]
    { 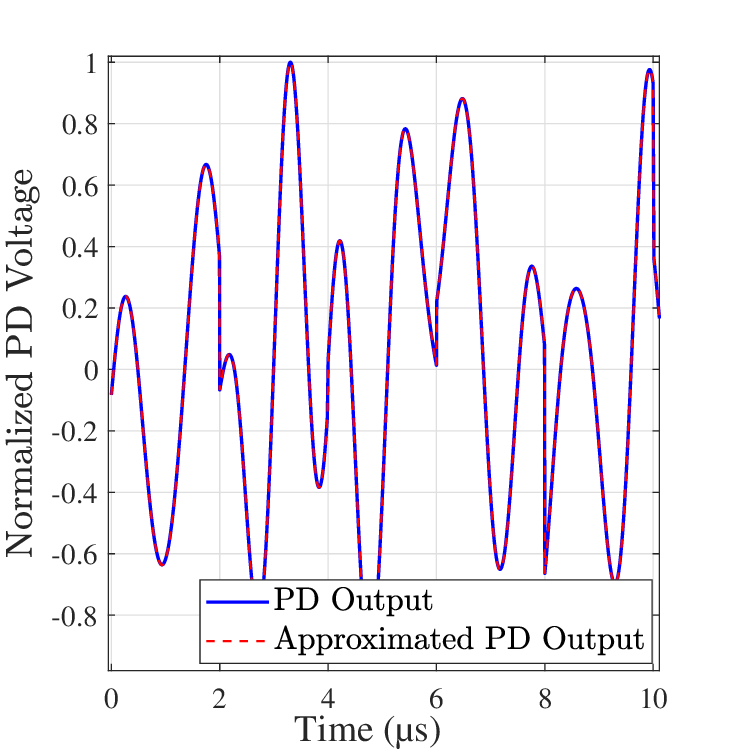}}
\subfloat[\centering] {\includegraphics[width=0.241 \textwidth]
    { 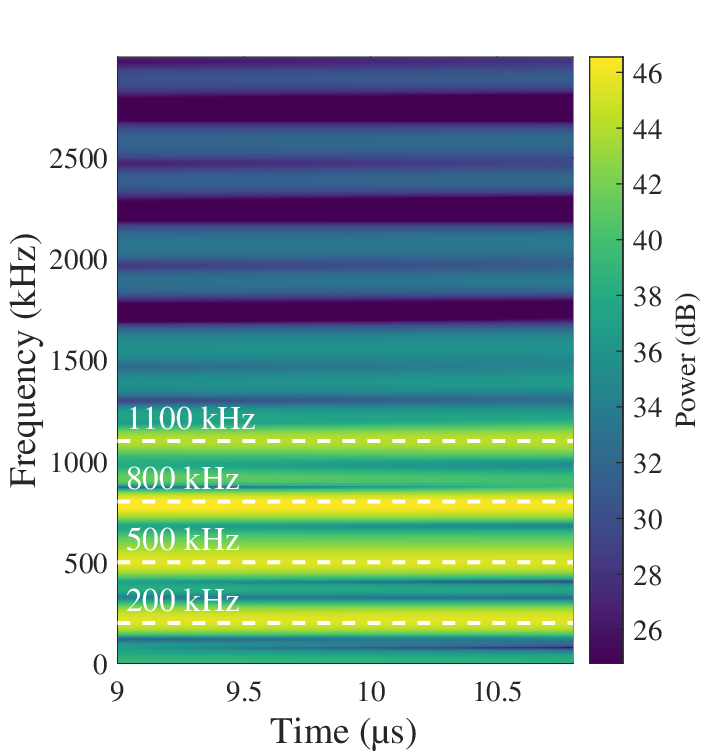}}
\caption{(a) Time-domain comparison between the DC-filtered and 
normalized exact photodetector output from 
\eqref{eq:photodetector_output} and the 
Taylor-approximated response $y(t)$ from 
\eqref{eq:linear_PD_output}. 
(b) Spectrogram of the DC-filtered photodetector output showing 
four clearly separated baseband channels at frequency offsets 
$\delta\omega_n$.}
\label{fig:taylor_pd}
\end{figure}

\subsection{Validation of the Linearized Photodetection Model}
\label{subsec:linear_validation}

We next validate the first-order Taylor approximation used to obtain the
RF-to-baseband signal model in Sec.~\ref{subsec:probe_transmission}.
Specifically, the exact photodetector output derived from the optical
susceptibility in \eqref{eq:photodetector_output} is compared with the linearized
response obtained from the Taylor expansion in \eqref{eq:linear_PD_output}.

Fig.~\ref{fig:taylor_pd}(a) shows the time-domain photodetector output.
The exact response and the Taylor-approximated signal are plotted after
DC removal and normalization to highlight the envelope dynamics.
The simulation is performed at the optimized operating point
$\boldsymbol{\Omega}^{\star}_{\mathrm{LO}}$ defined in
\eqref{OptimumRabiLO}. The two waveforms overlap almost perfectly,
demonstrating that the first-order Taylor approximation accurately
captures the probe-transmission response in the small-signal RF regime.
For this experiment, four independent RF signals are applied with carrier
offsets
$\delta\omega_1=200~\mathrm{kHz}$,
$\delta\omega_2=500~\mathrm{kHz}$,
$\delta\omega_3=800~\mathrm{kHz}$, and
$\delta\omega_4=1.1~\mathrm{MHz}$.
Each channel occupies a baseband bandwidth of $100~\mathrm{kHz}$ and
all signals are transmitted with equal power of $100~\mathrm{mW}$.
Additive noise is modeled as zero-mean Gaussian noise.
The spectrogram of the DC-filtered photodetector output is shown in
Fig.~\ref{fig:taylor_pd}(b). Four distinct spectral bands appear at the
corresponding offsets $\delta\omega_n$, confirming that the RF signals
are independently mapped to separable baseband components. The observed
frequency spread around each offset reflects the finite baseband
bandwidth of the transmitted signals.

These results confirm that the linearized photodetection model accurately
describes the RF-to-optical conversion process and enables simultaneous
extraction of multiple RF channels from a single probe transmission
measurement.

\begin{figure*}[t]
\centering
\subfloat[]{
    \includegraphics[width=0.3\textwidth]{ 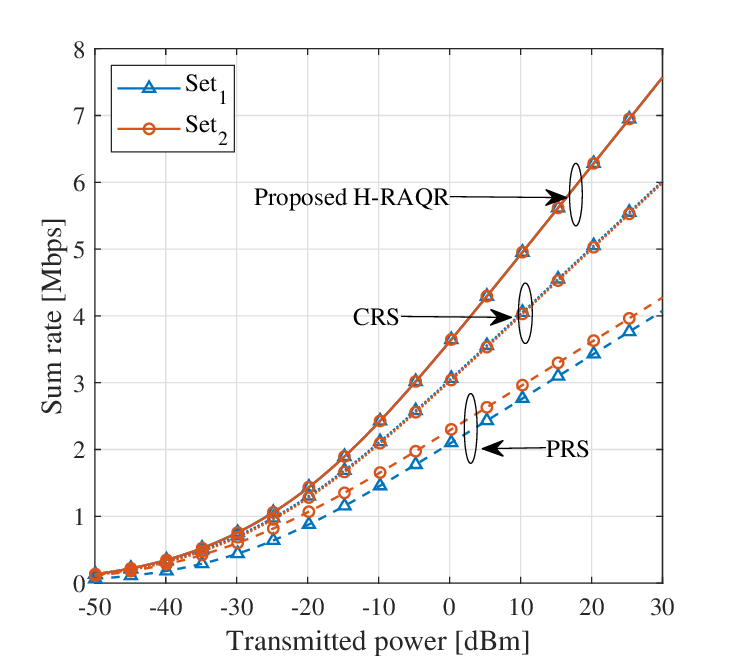}
}
\hfill
\subfloat[]{
    \includegraphics[width=0.3\textwidth]{ 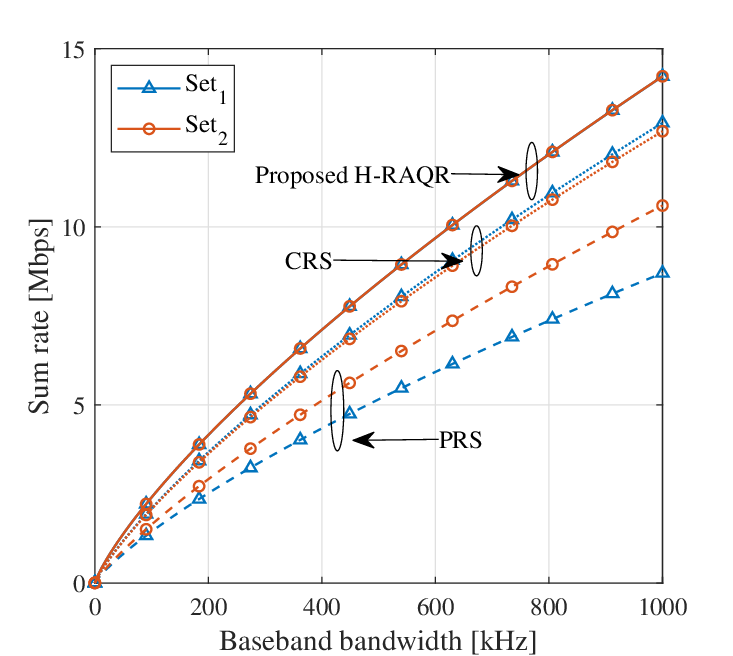}
}
\hfill
\subfloat[]{
    \includegraphics[width=0.3\textwidth]{ 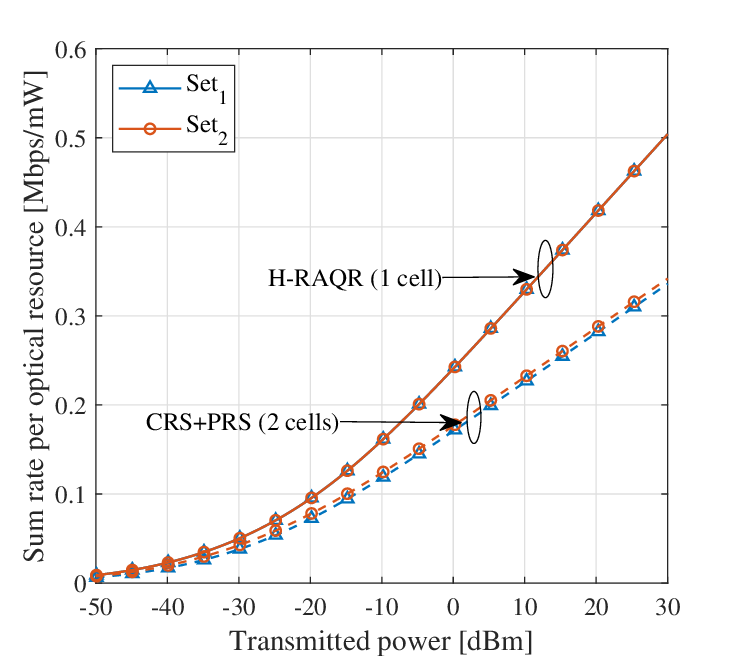}
}
\caption{(a) Achievable sum rate versus transmitted RF power for
$\mathrm{Set}_1$ and $\mathrm{Set}_2$ with baseband bandwidth
$B = 100$~kHz. (b) Rate versus baseband bandwidth for
$\mathrm{Set}_1$ and $\mathrm{Set}_2$ with transmitted power
$P_T = -10$~dBm. (c) Throughput per optical resource (Mbps/mW) versus transmitted RF power for $\mathrm{Set}_1$
and $\mathrm{Set}_2$, comparing the H-RAQR (one vapor cell)
against the combined CRS+PRS configuration (two vapor cells).}
\label{fig:performce}
\end{figure*}


\subsection{Communication Performance Comparison}
\label{subsec:rate_comparison}

We evaluate the communication performance of the proposed H-RAQR using the ergodic sum-rate metric
derived in Sec.~\ref{subsec:rate_analysis}. The performance is compared
against two reference architectures: the conventional CRS scheme and
the PRS scheme. All three receivers are simulated using the same
atomic system, vapor cell parameters, and optical laser configuration
to ensure a fair comparison.

Within the considered six-level manifold, the achievable number of RF
channels differs across the three architectures. The PRS scheme
supports only two RF transitions $(\omega_1,\omega_4)$ because any
additional intermediate coupling (e.g., $|3\rangle\leftrightarrow|5\rangle$)
would introduce a dipole-forbidden triangular loop as discussed in
Remark~1. The CRS architecture supports three sequential transitions
but does not activate the non-adjacent coupling
$|3\rangle\leftrightarrow|6\rangle$. In contrast, the proposed hybrid
architecture exploits the full dipole-allowed connectivity of the
six-level system and simultaneously activates four RF transitions.
Consequently, the hybrid receiver provides a larger number of 
communication channels within the same atomic structure. 
The achievable ergodic sum-rate results are shown in 
Fig.~\ref{fig:performce} present the communication performance for two
representative Rabi-frequency configurations (normalized by
$2\pi\times\mathrm{MHz}$),

\begin{equation}
\text{Set}_1=\{7,\,3,\,3,\,3\},
\qquad
\text{Set}_2=\{1,\,6,\,6,\,6\}.
\end{equation}
Both configurations lie within the validated operating region identified
in Section~\ref{sec:validityandLimitation}, achieving
$\mathcal{F}\approx0.98$ and therefore satisfying the applicability
conditions of the analytical model in \eqref{eq:rho21_steady}. The Rabi
frequencies are selected to ensure a fair comparison among the H-RAQR,
PRS, and CRS architectures.

Fig.~\ref{fig:performce}(a) shows the achievable rate as a function
of transmitted RF power with baseband bandwidth $B=100~\mathrm{kHz}$.
The hybrid receiver consistently achieves higher throughput across the
entire power range due to its ability to simultaneously utilize four RF
channels. Fig.~\ref{fig:performce}(b) shows achievable rate as a
function of baseband bandwidth with transmitted power fixed at
$-10~\mathrm{dBm}$, illustrating the effect of bandwidth allocation on
multichannel reception.
Across both scenarios and for both Rabi-frequency configurations, the
hybrid architecture outperforms the CRS-type and PRS-type receivers. For example, under $\text{Set}_2$ at a transmitted power of $-10~\mathrm{dBm}$ and $B=100~\mathrm{kHz}$, the proposed H-RAQR achieves approximately $2.39~\mathrm{Mbps}$. This corresponds to performance gains of roughly $16\%$ and $46\%$ over the CRS scheme ($2.06~\mathrm{Mbps}$) and PRS scheme ($1.63~\mathrm{Mbps}$), respectively.
These improvements arise from the hybrid coupling strategy, which
enables additional dipole-allowed RF transitions that are not accessible
in conventional CRS or PRS configurations within the same
six-level  atomic manifold.

\subsection{Resource-Efficiency Comparison}
\label{subsec:resource_efficiency_results}

We next evaluate the resource-efficiency metric in
\eqref{eq:resource_efficiency}. For an equivalent four-band reception
capability, the proposed H-RAQR employs a single vapor-cell front-end
($N_{\mathrm{FE}}=1$), whereas the conventional solution requires the
combined deployment of CRS and PRS receivers
($N_{\mathrm{FE}}=2$). The optical power
$P_{\mathrm{laser}}$ is computed using identical probe and coupling
Rabi frequencies for all architectures. Figure~\ref{fig:performce}(c) shows the resulting resource efficiency as
a function of the transmitted RF power for $\mathrm{Set}_1$ and
$\mathrm{Set}_2$. At $P_T=-10$~dBm, the proposed H-RAQR achieves
approximately $0.16~\mathrm{Mbps/mW}$, compared with
$0.124~\mathrm{Mbps/mW}$ for the combined CRS+PRS architecture,
corresponding to a gain of approximately $29\%$. This improvement stems
from the ability of the H-RAQR to support all four RF channels within a
single vapor-cell receiver, eliminating the additional optical
front-end required by the conventional architectures. Consequently, the
proposed receiver provides not only higher communication throughput
(Figs.~\ref{fig:performce}(a) and~\ref{fig:performce}(b)) but also
greater throughput per unit optical receiver resource.



\section{Conclusion}

This paper presented a hybrid six-level Rydberg atomic quantum receiver
(H-RAQR) for simultaneous multi-band RF reception within a single
vapor-cell receiver. By integrating parallel and cascaded RF coupling
pathways, the proposed architecture exploits the full dipole-allowed
connectivity of the Rydberg manifold to increase the number of
simultaneously accessible RF channels. A communication-oriented
analytical framework was developed by establishing the relationship
between atom--field interactions, the equivalent baseband signal model,
and the achievable communication performance. The analytical model was
validated against full Lindblad master-equation simulations within its
identified operating regime. Numerical results demonstrated that the
proposed H-RAQR achieves higher ergodic sum rate and greater resource
efficiency than conventional PRS and CRS receiver architectures while
providing equivalent four-band reception capability. These results
demonstrate the potential of hybrid Rydberg receivers as a scalable
platform for multi-band wireless communication. Future work will extend
the proposed framework to generalized multi-level hybrid architectures
supporting a larger number of simultaneous RF channels.


{\appendices
\section{Derivation of the Steady-State Density Matrix}
\label{app:density-matrix}

Under the resonant conditions in \eqref{eq:resonant_cond} and the dominant-decay approximation, retaining only $\gamma_{2,1}$. The steady-state density matrix $\boldsymbol\rho_{\mathrm{ss}}$ is obtained from the null space of the Liouvillian superoperator, $\boldsymbol{\mathcal L}\operatorname{vec}(\boldsymbol\rho_{\mathrm{ss}})=0$, subject to $\operatorname{Tr}(\boldsymbol\rho_{\mathrm{ss}})=1$, $\boldsymbol\rho_{\mathrm{ss}}=\boldsymbol\rho_{\mathrm{ss}}^\dagger$, and $\boldsymbol\rho_{\mathrm{ss}}\succeq0$. For compactness, we define

\begin{align}
\Sigma_{\Omega}^2&=\sum_{n=1}^{4}\Omega_n^2, \qquad \zeta=\Omega_1\Omega_3-\Omega_2\Omega_4,
\nonumber\\
\Lambda
&=
\zeta^2 \gamma_{2,1}^2
+2 \Omega_P^4 \Sigma_{\Omega}^2
+2\Big[(\Omega_2^2+\Omega_3^2)\Omega_C^2+\zeta^2\Big]\Omega_P^2.
\label{eq:app_sigma}
\end{align}

The steady-state density-matrix elements follow from the same elimination procedure and are listed below.

\begin{align}
\rho_{1,1} &= \frac{\zeta^2 \gamma_{2,1}^2 + \big[(\Omega_2^2 + \Omega_3^2)\Omega_C^2 + \zeta^2\big]\Omega_P^2}{\Lambda}, \label{eq:rho11} \\
\rho_{2,2} &= \frac{\Omega_P^2 \zeta^2}{\Lambda}, \quad \rho_{3,3} = \frac{\Omega_P^4 (\Omega_2^2 + \Omega_3^2)}{\Lambda},\label{eq:rho22} \\
\rho_{4,4} &= \frac{\Omega_P^2\big[\Omega_P^2(\Omega_3^2 + \Omega_4^2) + \Omega_C^2 \Omega_3^2\big]}{\Lambda}, \label{eq:rho44} \\
\rho_{5,5} &= \frac{\Omega_P^4 (\Omega_1^2 + \Omega_4^2)}{\Lambda}, \label{eq:rho55} \\
\rho_{6,6} &= \frac{\Omega_P^2\big[\Omega_P^2(\Omega_1^2 + \Omega_2^2) + \Omega_C^2 \Omega_2^2\big]}{\Lambda}. \label{eq:rho66}\\
\rho_{2,1} &= -\mathbbm{i}\,\frac{\Omega_P \zeta^2 \gamma_{2,1}}{\Lambda}, \quad \rho_{3,1} = -\frac{\Omega_P^3 \Omega_C (\Omega_2^2 + \Omega_3^2)}{\Lambda}, \label{eq:rho21_app}
\end{align}
\begin{align}
\rho_{4,1} &= \mathbbm{i}\,\frac{\Omega_P \Omega_C \Omega_3 \zeta \gamma_{2,1}}{\Lambda}, \quad \rho_{4,2} = -\frac{\Omega_P^2 \Omega_C \Omega_3 \zeta}{\Lambda},  \label{eq:rho41_app} \\
\rho_{5,1} &= \frac{\Omega_P^3 \Omega_C (\Omega_1 \Omega_2 + \Omega_3 \Omega_4)}{\Lambda}, \label{eq:rho51_app} \\
\rho_{6,1} &= -\mathbbm{i}\,\frac{\Omega_P \Omega_C \Omega_2 \zeta \gamma_{2,1}}{\Lambda}, \quad \rho_{6,2} = \frac{\Omega_P^2 \Omega_C \Omega_2 \zeta}{\Lambda},      \label{eq:rho61_app} \\
\rho_{5,3} &= -\frac{\Omega_P^4 (\Omega_1 \Omega_2 + \Omega_3 \Omega_4)}{\Lambda}, \label{eq:rho53_app} \\
\rho_{6,4} &= -\frac{\Omega_P^2\big[\Omega_P^2(\Omega_2 \Omega_3 + \Omega_1 \Omega_4) + \Omega_C^2 \Omega_2 \Omega_3\big]}{\Lambda}. \label{eq:rho64_app}
\end{align}
The remaining lower-triangular coherences vanish, i.e.,
$\rho_{3,2},\rho_{4,3},\rho_{5,2},\rho_{6,3},\rho_{5,4},\rho_{6,5}=0$, while upper-triangular elements follow from Hermiticity, $\rho_{j,i}=\rho_{i,j}^{*}$ for $i>j$.




\section{Extended Fidelity Maps}
\label{app:fidelity}

\begin{figure}[H]
    \centering
    \includegraphics[width=0.9\linewidth]{ 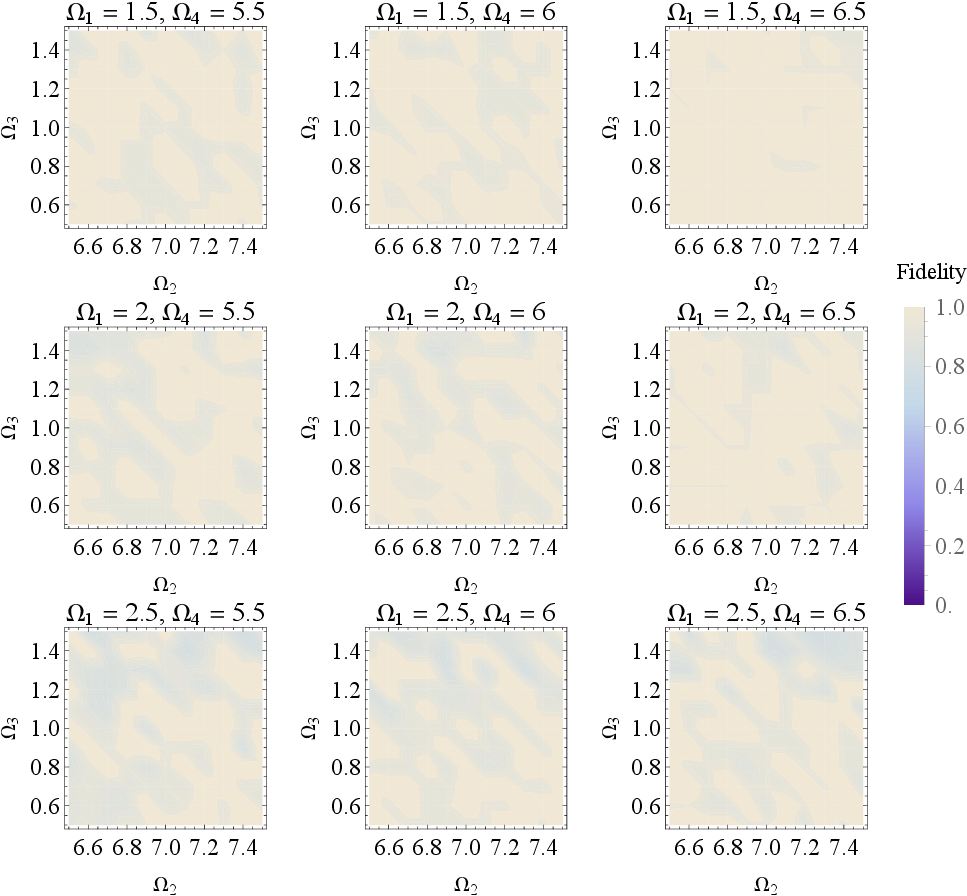}
     \caption{Zoomed fidelity landscape around selected LO operating point 
$(\Omega_1,\Omega_2,\Omega_3,\Omega_4)
=
(2\pi\times2,\,2\pi\times7,\,2\pi\times1,\,2\pi\times6)\,\mathrm{MHz}$. A pronounced high-fidelity plateau is observed, confirming 
the robustness of the analytical steady-state model to small RF-induced 
perturbations. 
}
    \label{fig:fidelityAppendix}
\end{figure}}


\end{document}